\documentclass[%
 aip,
 amsmath,amssymb,
 reprint,%
]{revtex4-1}

\pdfoutput=1
\usepackage[english]{babel}
\usepackage{graphicx,bm,amsmath,eufrak,amssymb,url}
\usepackage[usenames,dvipsnames]{color}
\usepackage[bbgreekl]{mathbbol}
\usepackage{xcolor}
\usepackage{placeins}
\usepackage{algcompatible}
\usepackage{algorithm}
\usepackage{algpseudocode}
\usepackage{braket}
\usepackage[colorlinks=true, breaklinks=true, linkcolor=darkblue, citecolor=darkblue, urlcolor=darkblue]{hyperref}
\usepackage{pgf}
\usepackage[T1]{fontenc}
\usepackage{verbatim}

\usepackage{graphicx}
\usepackage{amsthm}
\usepackage{hyperref}
\usepackage[font=small,labelfont=it,justification=justified,format=plain]{caption}
\usepackage{subcaption}
\usepackage{url}
\usepackage{ulem}
\usepackage{cancel}
\usepackage{amsmath}

\DeclareMathOperator*{\argmin}{arg\,min}
\DeclareMathOperator{\diag}{\mathrm{diag}}

\renewcommand{\H}{\hat{H}}

\newcommand{\tn}{\bm{t}^{(n)}}
\newcommand{\tnn}{\bm{t}^{(n+1)}}
\newcommand{\U}{\hat{U}}
\newcommand{\Ubar}{\hat{\mathcal{U}}}
\newcommand{\kap}{\hat{\kappa}}

\definecolor{darkblue}{rgb}{0, 0, 0.8}

\newtheorem{theorem}{Theorem}
\newtheorem{lemma}{Lemma}

\usepackage{environ}
\NewEnviron{align2}{
\begin{subequations} \begin{align} \BODY \end{align} \end{subequations}
            }

\counterwithin{equation}{subsection}
\begin{document}
\title{Optimizing unitary coupled cluster wave functions on quantum hardware: error bound and resource-efficient optimizer}

\author{Martin Plazanet}
\affiliation{Eviden Quantum Lab, Les Clayes-sous-Bois, France}
\affiliation{Centre for Quantum Technologies, National University of Singapore, Singapore 117543, Singapore}
\affiliation{École Polytechnique, I.P. Paris, F-91120 Palaiseau, France}
\affiliation{MajuLab, CNRS–UCA-SU-NUS-NTU International Joint Research Laboratory}
\author{Thomas Ayral}
\affiliation{Eviden Quantum Lab, Les Clayes-sous-Bois, France}
\begin{abstract} 
   
   In this work, we study the projective quantum eigensolver (PQE) approach to optimizing unitary coupled cluster wave functions on quantum hardware, as introduced in Ref. \onlinecite{Stair_2021}. The projective quantum eigensolver is a hybrid quantum-classical algorithm which, by optimizing a unitary coupled cluster wave function, aims at computing the ground state of many-body systems. Instead of trying to minimize the energy of the system like the variational quantum eigensolver, PQE uses projections of the Schrödinger equation to  efficiently bring the trial state closer to an eigenstate of the Hamiltonian. 
   In this work, we provide a mathematical study of the algorithm. 
   We derive a bound relating off-diagonal coefficients (residues) of the Hamiltonian to the energy error of the algorithm and the overlap achieved by the obtained wavefunction. These bounds not only give formal guarantees to
   PQE, but they also allow us to formulate a well-informed convergence criterion for residue-based optimizers.
   We then study the classical optimization itself and derive convergence guarantees under certain conditions.  
We propose a new residue-based optimizer,
 with numerical evidence of the superiority of this new approach for  H$_4$, H$_6$, BeH$_2$ and LiH dissociation curves over both the optimization introduced in Ref.~\onlinecite{Stair_2021} and VQE optimized using the Broyden–Fletcher–Goldfarb–Shanno (BFGS) method.
\end{abstract}
\maketitle

Simulating quantum many-body physics is the prime motivation behind the inception of quantum computing \cite{Feynman1982}.
The main premise of quantum computers is that a time evolution of these artificial quantum systems \cite{Lloyd1996}---with the laws of quantum mechanics built-in---will not be plagued with the computational bottlenecks attached to mimicking these dynamics with classical computers \cite{Ayral2023}.
Yet, the experimental quantum processors that have appeared in the recent years must reckon with decoherence, an unavoidable byproduct of the very laws that afford a quantum acceleration.
With quantum error correction still a long way ahead despite very promising recent experimental efforts \cite{Krinner2021, Bluvstein2023, DaSilva2024, Acharya2024}, textbook quantum algorithms like quantum phase estimation \cite{kitaev1995quantum}, whose main ingredient is the aforementioned time evolution, are rendered virtually useless by decoherence because of the duration, often measured in terms of the gate count, of the corresponding quantum circuit.

This practical hurdle has prompted the revival of the standard variational method in a quantum context: while the quantum processor is used to prepare a possibly complex parametric quantum state and to measure the corresponding variational energy, the new parameters are determined by a classical optimization algorithm.
This hybrid algorithm, called the variational quantum eigensolver (VQE, \cite{Peruzzo_2014}, see Ref.~\onlinecite{Tilly_2022} for a review), a priori allows one to pick a variational state whose preparation on the quantum processor will be faster than decoherence, while being complex enough that its energy would be hard to measure on a classical processor alone.
However, recent research efforts have unveiled a number of issues with VQE.
(i) {\it Decoherence}: it sharply limits the size of circuits that can be run in practice and therefore their expressivity. Beyond compilation efforts to shorten a given state preparation with sophisticated optimizations of the circuit at hand, adaptive ansatz construction methods \cite{Grimsley2019} or orbital optimization methods \cite{Mizukami2020, Besserve2021, Ratini2023, Moreno2023, Besserve2024} have been developed to make the most of the available coherence. But even without decoherence, VQE would still suffer from
(ii) {\it the measurement problem}: measuring the variational energy involves computing the empirical means of quantum measurement outcomes, yielding a central-limit-theorem-induced $1/\epsilon^2$ dependence of the run time on the statistical error $\epsilon$, leading to prohibitively long run time estimates for relevant quantum chemistry problems \cite{Wecker_2015}.
Improvements over plain vanilla energy estimation like term grouping \cite[Section~5]{Tilly_2022} or classical shadows \cite{Huang2020} may improve the prefactor but do not alter this $1/\epsilon^2$ dependence. 
Lastly, VQE can be plagued with (iii) {\it an optimization problem}, namely by  the hardness of finding the global minimum of the variational energy.
Very long, and thus potentially very expressive, state preparation circuits like so-called hardware-efficient ansätze \cite{Kandala2017} tend to be close to random, which leads to a cost landscape that is exponentially flat with the number of orbitals (qubits) \cite{Ragone2023, Larocca2024}.
This "barren plateau" phenomenon \cite{McClean_2018,Larocca2024} is but an average statement, and could therefore be escaped with carefully initialized, and/or physically motivated variational states.
Yet, even in the absence of this problem, the precise behavior of the optimization is a still largely uncharted territory.
In particular, the number of steps to convergence and the distance to the true ground state energy are in general unknown in VQE.

Facing these issues, two main routes have been taken thus far.
A first route consists in turning back to quantum phase estimation algorithms and, waiting for quantum error corrected computers, optimizing the various building blocks of the algorithm to reduce the requisite resources.
This is arguably a very long term strategy given the chasm between the number of physical qubits required for such a quantum error corrected approach \cite{Wecker2014, Beverland2022}, and the number of physical qubits achieved in recent experiments \cite{Acharya2024}.
Also, a prerequisite for an efficient quantum phase estimation is an input state with a large enough overlap with the sought-after ground state. Finding such an input state is probably difficult for very correlated systems \cite{Lee2022, Louvet2023}.
This underlines the importance of the second route, which consists in devising ground state preparation strategies suited for current or near-term processors.
This route relies on the hope either that quantum advantage can be squeezed out of these processors in the near term, or that such methods will anyway be needed to produce good enough input states for error-corrected algorithms, implemented on fault-tolerant processors, like phase estimation.

Recent representatives of this second route include recent works like that of Refs~\onlinecite{Robledo-Moreno2024, Yu2025} that supposedly eschew the optimization and measurement problems by not performing any optimization on the quantum computer (the optimal parameters found by a classical method are fed to a quantum ansatz without reoptimization) and by not attempting the estimation of the energy on the quantum computer, but instead use the quantum processor to select important subspaces, which a classical supercomputer then diagonalizes.

Another method, which will be the focus of this work, turns away from the variational principle and instead, taking direct inspiration from the numerical solution of coupled cluster equations \cite{10.1063/1.1727484} on classical computers, formulates ground state search as a root-finding problem.
This method, dubbed the projective quantum eigensolver (PQE), was introduced in Ref.~\onlinecite{Stair_2021}.
Its superiority in terms of speed compared to VQE was checked on a few example systems using numerical simulations. Ref.~\onlinecite{Stair_2021} also mentions the existence of an error bound.
The algorithm was also implemented recently \cite{Misiewicz2023} on superconducting hardware.

In this paper, we build on the promising properties of PQE, namely the faster convergence observed in the numerical benchmarks, and the purported existence of error bounds, which indicate the algorithm
could possibly
not suffer from phenomena such as local minima. 
Here, we provide rigorous and tighter error bounds for PQE.
We then introduce a mathematical study of quasi-Newton methods for PQE, ultimately allowing us to introduce a novel optimization method with a faster and more robust convergence.
We provide extensive numerical benchmarks showing the superiority in resource cost of this method over both VQE optimized using the Broyden–Fletcher–Goldfarb–Shanno (BFGS) method, and the original optimization for PQE proposed in Ref.~\onlinecite{Stair_2021}.

The paper is structured as follows:
In the first section we summarize the projective quantum eigensolver approach as introduced in Ref.~\onlinecite{Stair_2021} as well as our main results. 
In the second section we introduce an error bound for the algorithm, and discuss its implications and utility.
In a third section, we introduce a mathematical study of quasi-Newton algorithms and discuss its implications. 
In the last part, we introduce our own, more robust and resource-efficient optimization method.

\section{Summary of the projective quantum eigensolver method and of our results} 
\subsection{The Algorithm}
\label{brief_summary}

In this section we briefly summarize the projective quantum eigensolver method introduced by Ref.~\onlinecite{Stair_2021}.

We are considering a system of $m$ interacting fermionic modes described by the electronic structure Hamiltonian: 
\begin{equation} \label{hamiltonian}
\hat{H} = \sum_{pq=1}^m h_{pq}\hat{a}_p^\dag  \hat{a}_q + \frac12 \sum_{pqrs=1}^m v_{pqrs} \hat{a}_p^ \dag \hat{a}_q^\dag \hat{a}_s \hat{a}_r, 
\end{equation}
where  $h_{pq}$ is a one-body integral, $v_{pqrs}$ a two-body integral, and $\hat{a}_p^ \dag$ and $\hat{a}_q$ respectively create and annihilate electrons in the corresponding spin-orbitals. 
We wish to compute this Hamiltonian's ground state energy $E_0$ by optimizing a parametric representation
$\ket{\Psi(\bm{t})} = \hat{U}(\bm{t})\ket{\Phi_0}$ of the many-body wavefunction, with $\ket{\Phi_0}$ a reference state (usually taken to be the Hartree-Fock state), and $\U(\bm{t})$ a unitary parameterized by parameters $\bm{t}\in \mathbb{R}^N$.

The number $N$ of parameters is assumed to be much smaller than the Hilbert space size, $N\ll 2^m$, to benefit from a compressed (not exponentially costly) representation of the wavefunction and to enable (although not guarantee) an efficient optimization process.
Here, as in Ref.~\onlinecite{Stair_2021} and in many quantum chemical approaches, we shall use a variant of the unitary coupled cluster ansatz often referred to as 'disentangled' \cite{Evangelista_2019}:
\begin{equation} \label{ansatz}
\U(\bm{t}) = \prod_{\mu\in \mathcal{A}} e^{t_\mu \hat{\kappa}_\mu},
\end{equation}
where the generators $\hat{\kappa}_\mu$ read $\kappa_\mu = \hat{\tau_\mu} - \hat{\tau}^\dag_\mu$, with excitation operators $\tau_{(i,j, \cdots) (a,b, \cdots)} = \hat{a}_a^ \dag \hat{a}_b^\dag \cdots \hat{a}_j \hat{a}_i $,  representing an excitation from occupied orbitals $i,j, \dots$ to unoccupied orbitals $a,  b, \dots$.
These generators are chosen from a pool called $\mathcal{A}$, 
which itself is a matter of choice or chemical intuition. 
In practice we will usually limit ourselves to single and double excitations, and sometimes also triple excitations.
It is also possible, by taking inspiration from adaptive variants of VQE \cite{Grimsley2019}, to pick the most relevant excitations adaptively. This is described by, and used in, Ref.~\onlinecite{Stair_2021}.

Contrary to variational approaches like VQE, which aim at minimizing the parametric energy 
\begin{equation}
    E(\bm{t}) = \bra{\Psi(\bm{t})} \H \ket{\Psi(\bm{t})}\label{eq:parametric_energy},
\end{equation}
the PQE method consists in optimizing the parameters $\bm{t}$ to bring our state as close to an eigenstate as possible. 
This is achieved by considering an orthonormal basis $\{ \ket{\Phi_\nu}\}$ containing $\ket{\Phi_0}$ (which, when $\ket{\Phi_0}$ is the Hartree Fock (HF) state, will correspond to excited HF determinants), and by minimizing the so-called residues:
\begin{equation} \label{def_r_mu}
r_\nu(\bm{t})= \bra{\Phi_\nu}\hat{U}^\dag(\bm{t})\hat{H}\hat{U}(\bm{t})\ket{\Phi_0}.
\end{equation}
These residues correspond to the projection of $\hat{H}\ket{\Psi(\bm{t})}$ onto our  basis (rotated by $U(\bm{t})$). 
PQE thus boils down to solving the nonlinear system of equations:
\begin{align} \label{pqe_equations}
\forall \nu \neq 0, \ r_\nu(\bm{t}) &= 0,
\end{align}
for $\bm{t}\in\mathbb{R}^N$.
This set of equations is very similar to the set of equations that is solved in the coupled cluster method in its projective (as opposed to variational) variant \cite{RevModPhys.79.291}. 
The main difference between PQE and projective coupled cluster is that PQE uses a {\it unitary} coupled cluster ansatz, which is efficiently implementable as a quantum circuit.

\begin{figure}[t]
    \centering
    \includegraphics[width = 1.0\columnwidth]{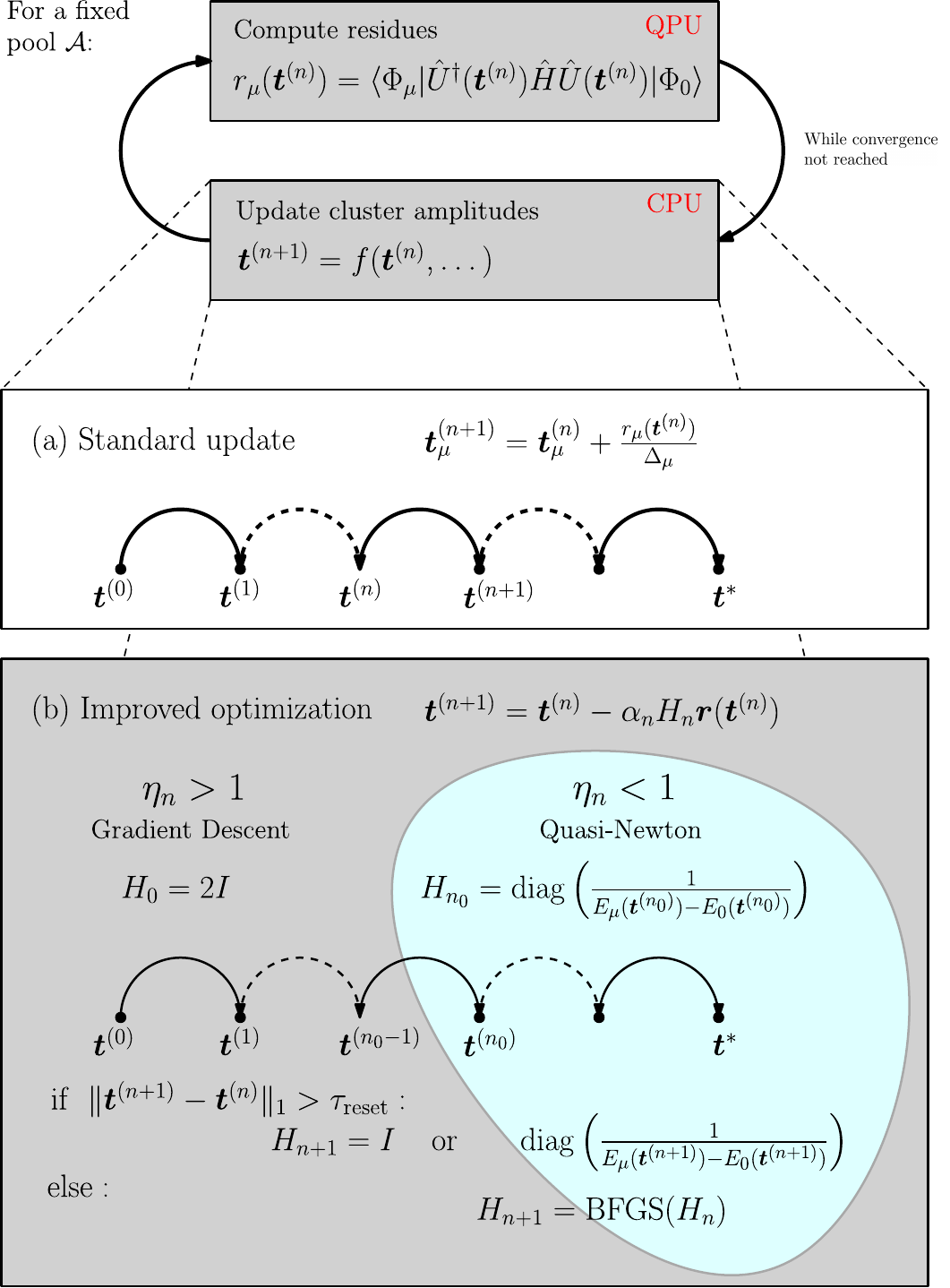} 
    \caption{Outline of the projective quantum eigensolver for a fixed operator pool $\mathcal{A}$. The residues $r_\mu$ are computed using Eq.~\eqref{compute_r_mu}. Two update rules for the cluster amplitudes $\bm{t}$ are depicted: (a) Standard update proposed in \cite{Stair_2021}, with $\Delta_\mu$ defined in Eq.~\eqref{eq:moller_plesset}. (b) Improved update (introduced in section \ref{sec:Better_opti}, Algorithm \ref{pseudocode}), which (mainly) depends on the value of $\eta_n$ (Eq.~\eqref{eq:eta_def}).} 
    \label{fig:sketch}
\end{figure}

At this stage, two practical implementation remarks are in order:
First, since the number of possible excited states $\ket{\Phi_\nu}$ grows exponentially with the number of orbitals, in practice, one solves the subsystem on a subset of the possible excitations. Here, we will choose the same---possibly adaptively chosen---subset $\mathcal{A}$ as the one used to generate the ansatz. We will therefore use the same indices $\mu$ for the residues as for the parameters. The mathematical motivations of this choice will become more apparent in section \ref{math_section}.

Second, in practice, the solution of this nonlinear system of equations is nontrivial. Ref.~\onlinecite{Stair_2021} advocates an approximation of a modified Newton-Raphson process. Instead of the standard Newton-Raphson strategy, which updates parameter $\bm{t}^{(n)}$ at step $n$ following the update rule
\begin{equation} \label{newton_update_rule}
\bm{t}^{(n+1)} = \bm{t}^{(n)} - \bm{J}^{-1}(\bm{t}^{(n)})\bm{r}(\bm{t}^{(n)}),
\end{equation}
the \textit{modified} Newton-Raphson method replaces $\bm{J}^{-1}(\bm{t}^{(n)})$ with $\bm{J}^{-1}(\bm{t}^{(0)})$ to avoid the cost of computing and inverting the Jacobian matrix $J_{\mu \mu'} (\bm{t}) =  \partial r_\mu / \partial t_{\mu'} (\bm{t}) $ at every step.
(In the case of a Hartree Fock starting point, we will have $\bm{t}^{(0)} = 0$). 
After various approximations which will be elaborated on in a later section, Ref.~\onlinecite{Stair_2021} further simplifies the update rule to
\begin{equation} \label{standard-update-rule}
    \tnn_\mu = \tn_\mu + \frac{r_\mu(\tn)}{\Delta_\mu}, 
\end{equation} 
where
\begin{equation}
  \Delta_\mu = \Delta_{ij\cdots}^{ab\cdots} = \epsilon_i + \epsilon_j + \cdots - \epsilon_a - \epsilon_b - \cdots 
  \label{eq:moller_plesset}
\end{equation}
is the M\o ller-Plesset denominator ($\epsilon_i$ is the Hartree-Fock energy corresponding to the $i$th orbital). 
Fig.~\ref{fig:sketch} illustrates the PQE algorithm. The standard update proposed by Ref.~\onlinecite{Stair_2021} is depicted in box (a).

Each update step requires the evaluation of the residue $r_\mu$. As shown in Ref.~\onlinecite{Stair_2021}, it can be expressed, using the properties $\hat{\kappa}_\mu \ket{\Phi_0}= \ket{\Phi_\mu}$, $\hat{\kappa}_\mu^2 \ket{\Phi_0} = - \ket{\Phi_0}$, and the ensuing relation $e^{t_\mu \hat{\kappa}_\mu}\ket{\Phi_0} = \cos(t_\mu)\ket{\Phi_0} + \sin(t_\mu)\ket{\Phi_\mu}$, as
\begin{equation} \label{compute_r_mu}
r_\mu (\bm{t}) = E^{\frac{\pi}4}_\mu (\bm{t}) - \frac12 E_\mu (\bm{t}) - \frac12 E (\bm{t}),
\end{equation}
where we defined:
\begin{align}
    E^{\frac{\pi}4}_\mu (\bm{t}) &= \bra{\Phi_0} e^{-\frac{\pi}4 \hat{\kappa}_\mu} \hat{U}(\bm{t})^\dag \hat{H} \hat{U}(\bm{t}) e^{\frac{\pi}4\hat{\kappa}_\mu} \ket{\Phi_0},\\
    E_\mu(\bm{t}) &= \bra{\Phi_\mu}\hat{U}(\bm{t})^\dag \hat{H} \hat{U}(\bm{t}) \ket{\Phi_\mu}. \label{eq:E_mu}
\end{align}
This means each optimization step  requires only $2N+1$ measurements, where $N$ is the size of the operator pool. This is roughly the same cost as computing a gradient via a parameter-shift rule in VQE \cite{Schuld_2019,D0SC06627C}.
Note that one could obtain the residues using some other measurements of the same type (see for instance Ref.~\onlinecite[Equation 7]{misiewicz2023implementationprojectivequantumeigensolver}). 

Such a procedure was shown, in Ref.~\onlinecite{Stair_2021}, to yield better convergence properties than VQE for hydrogen chains (until 10 atoms), the algorithm yielding a similar accuracy when using the same ansatz, while requiring fewer optimization steps.

In the next sections, we will introduce an error bound for the algorithm, and then focus on a mathematical study of the optimization process in order to improve it.

\subsection{Summary of our results}
In this work, we mainly introduce two new results.

First, in section \ref{absolute_bound}, we introduce an error bound on the energy that is returned by the PQE algorithm.
This endows PQE with accuracy guarantees which VQE lacks: while a small gradient norm does not guarantee any proximity to the target energy (which is why problems such as local minima and barren plateaus can arise in VQE), a small residual norm does. 

\begin{figure}[H]
    \centering
    \includegraphics[width=1\linewidth]{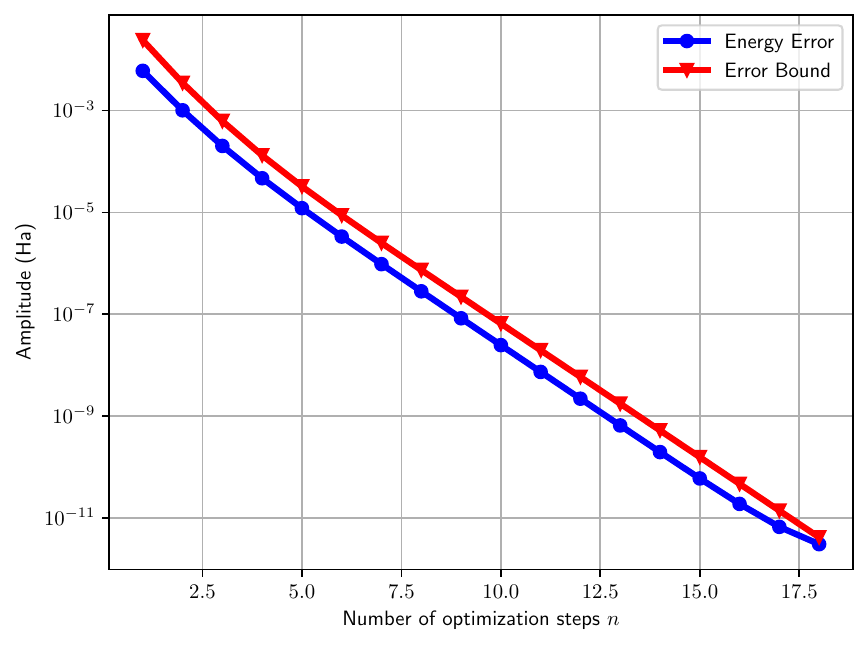}
    \caption{Test of our bound by running the PQE algorithm on an $\mathrm{H_4}$ molecule at $d_{H-H} = 0.8\text{\AA}$ with a complete UCCSDTQ ansatz and using a convergence threshold $\epsilon = 10^{-10}$. The blue slope represents the true energy error, {\it i.e} the difference $\varepsilon_{\mathrm{exact}}(\tn)$  (defined in Eq. \eqref{hf_error_bound} below) between the running energy and the exact ground state energy (calculated using FCI). The red slope represents the bound on $\varepsilon_\mathrm{exact}$ we derive in this work.} 
    \label{bound_showcase}
\end{figure}

The tightness of our bound is illustrated in Figure \ref{bound_showcase} for a $\mathrm{H}_4$ molecule. In this example, our bound (red curve) is systematically within less than of an order of magnitude of the true energy error (blue curve).
We discuss, in the sections below, the usefulness of this bound in terms of the convergence properties of the algorithm. We also use it to design a better stopping criterion.

\begin{figure*}[t]
    \centering
    \includegraphics[width=1\textwidth]{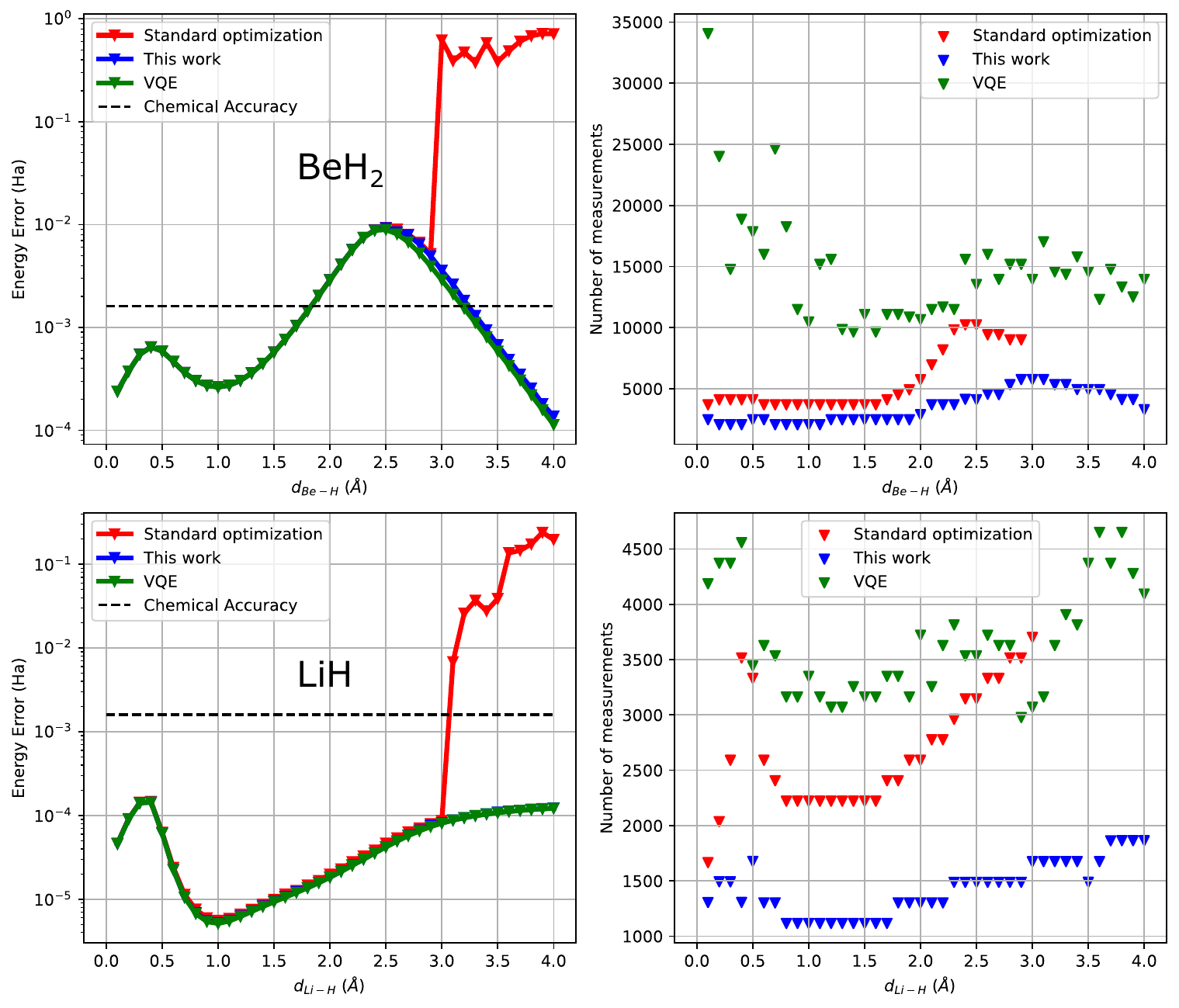}
    \caption{Comparison between the standard optimization of PQE (in red), our optimization (in blue) and VQE optimized using BFGS (in green) for computing the ground state energy of a $\mathrm{BeH_2}$ (top) and LiH (bottom) molecule using a complete UCCSD ansatz at different bond distances. Left panels: energy error, i.e the difference $\varepsilon_\mathrm{exact}$ (see Eq.~\eqref{hf_error_bound} below) between the energy value yielded by the algorithm and the exact FCI ground state energy computed. Right panels: total number of energy measurements (ie $\bra{\Psi}\H\ket{\Psi}$ for some state $\ket{\Psi}$) performed throughout the algorithm. The $x$ axis of each plot represents the bond distance.}
    \label{BeH2_vqe_sn_truc_1}
\end{figure*} 
Our second contribution, introduced in section \ref{sec:Better_opti}, is an optimization algorithm tailored for PQE.
It switches between Newton-Raphson updates and gradient-descent-like updates based on mathematically-grounded criteria: essentially, far from the solution, updates resembling gradients descent are chosen, while approximate Newton-Raphson updates are used when closing in on the solution.
This is illustrated in Fig.~\ref{fig:sketch}, box (b), which sketches the original PQE update rule as well as the main ideas behind our improved update rule.

A performance benchmark of our new method versus VQE and the original PQE algorithm (``Standard optimization'') is presented in Fig. \ref{BeH2_vqe_sn_truc_1}. For each of the three methods, it compares the energy error (left panels) and total number of measurements for different bond distances of the $\mathrm{BeH}_2$ and $\mathrm{LiH}$ molecules.
Two interesting features of our method stand out from this numerical comparison: it is more robust than the original PQE update rule for large bond lengths (where static correlations are larger), and much more economical than VQE (and standard PQE) in terms of the number of measurements (and hence overall run time) needed to reach a similar accuracy.

\section{A general error bound for the projective quantum eigensolver} \label{absolute_bound}
The projective quantum eigensolver computes an approximation of the ground state by using a simple rationale: if all the residues are zero, we are guaranteed to have an eigenstate of the Hamiltonian. Ref.~\onlinecite{Evangelista_2019} shows that any state, and thus the exact ground state, can be exactly represented by using a complete disentangled Unitary Coupled Cluster ansatz.

In practice, however, we will be looking at a portion of the residues, and at the end of the algorithm those will not be strictly zero, but rather smaller than a chosen threshold. Whether such a threshold can be reached will be discussed semi-locally in section \ref{math_section}.  It is therefore natural to wonder what guarantees on the accuracy of the result can be had by reducing residues. Moreover, the existence of such guarantees would mean that small residues imply a small error on the energy, which is not a property energy gradients in VQE have ($\vec{\nabla} E \approx  0$ does not imply that the distance to the ground state energy vanishes). 

Ref.~\onlinecite{Stair_2021} briefly mentions a bound on the error based on Gershgorin's circle theorem. It is however rather loose in general, and the proof is only valid if:
\begin{align}
    \max_\mu(|\bra{\Phi_\mu(t)}\Psi\rangle|^2)<|\bra{\Phi_0(t)}\Psi\rangle |^2,
\end{align}
where $\ket{\Psi}$ is the ground state. This is not valid in highly correlated systems when starting from the Hartree Fock state.

In this section, we introduce a general bound on the error, which stems from Kato's inequality. Kato's inequality has the upside of being very flexible  as to the assumptions we make, meaning it can be used in order to derive error bounds under rather weak assumptions, as will be the focus of subsection \ref{bound_special_case}, or under stronger assumptions to obtain tighter error bounds, as will be the focus of subsection \ref{temple-sec}.
We also introduce a simple lower bound on the overlap between the trial state and an eigenstate, which is a highly relevant quantity if one wishes to use PQE as a starting point for quantum phase estimation. 
We first introduce the mathematical results in subsection \ref{math-preliminaries}, and then discuss their application to the projective quantum eigensolver in subsection \ref{application_to_pqe}.

\subsection{Mathematical preliminaries} \label{math-preliminaries}

In this section we state existing theorems that allow to lower- and upper-bound the eigenvalue $E$ associated with the eigenvector $|\Psi\rangle$ of a Hamiltonian $\mathcal{H}$ and lower bound its overlap with a given trial state $|\psi\rangle$, given the energy $\mathcal{E}$ and variance $\mathcal{R}^2$ of the trial state.

\subsubsection{Bounding the eigenvalue}
We begin with the introduction of the most general bound, Kato's inequality.
\begin{theorem}{Kato's Inequality \cite[Theorem 2]{harrell1978generalizations}} \label{Kato_inequality} \\
    Let $\mathcal{H}$ be a self-adjoint operator on a Hilbert space with sesquilinear inner  product $\braket{\cdot | \cdot}$.
    Let $\ket{\psi}$ be a normalized vector of our Hilbert space. Define the energy $\mathcal E \equiv \bra{\psi}\mathcal{H}\ket{\psi} $ and the variance $\mathcal R^2 \equiv\bra{\psi}\mathcal{H}^2\ket{\psi}  - \bra{\psi}\mathcal{H}\ket{\psi}^2 $. Assume that there exist $\alpha, \ \beta \in \mathbb R$ such that $\alpha < \beta$ and:
    \begin{align}  \label{kato_inequality_1}
        \mathcal R^2 < \left(\beta -\mathcal E\right)\left( \mathcal E - \alpha\right).
    \end{align}
    Then there is at least one eigenvalue of $\mathcal H$ in $\left]\alpha, \beta \right[$. If there is only one such eigenvalue $E$, then we have the following inequality:
    
    \begin{align} \label{kato_inequality_2}
         \mathcal E - \frac{ \mathcal R^2}{\beta  -\mathcal E} \leq E \leq \mathcal E + \frac{ \mathcal R^2}{\mathcal E-\alpha} 
    \end{align}
    
    \end{theorem}
Looking the inequality, we can see that the interesting part is Eq. \eqref{kato_inequality_2}: it gives us an error bound on an eigenvalue of $\mathcal H$. However, the conditions for its applicability can seem rather unpractical. Although finding $\alpha, \beta$ such that Inequ. $\eqref{kato_inequality_1}$ holds is easy, finding them such that there is only one eigenvalue in  $\left]\alpha, \beta \right[$ complicates things. We can however see that if $\mathcal H$ has a spectral gap $\Delta$, then $\beta - \alpha < \Delta$ suffices to make sure that there is at most one eigenvalue in $\left]\alpha, \beta \right[$. Looking again at $\eqref{kato_inequality_1}$, we can see that:
\begin{align} \label{R-spectral-gap}
    \mathcal R^2 < \frac{\Delta^2}{4}
\end{align}
is a sufficient condition to ensure there exist $\alpha, \beta$ such that inequality \eqref{kato_inequality_2} holds. As we will later see in this section, $\mathcal R$ relates to the residual norm during PQE, and thus inequalities such as \eqref{kato_inequality_2} are particularly helpful in the context of PQE calculations.

\subsubsection{A lower bound on the overlap}
In the next subsection, we introduce a simple lower bound on the overlap between the trial state and an eigenstate.
We begin with Lemma $4$ from Ref. ~\onlinecite{harrell1978generalizations}.
\begin{lemma}{Ref.~\onlinecite{harrell1978generalizations}} \label{lemma_overlap}
    Let $E$ be an eigenvalue of $\mathcal H$, isolated from the rest of the spectrum by a distance $\Delta$, with corresponding eigenvector $\ket \Psi$. Let $\ket \psi$ such that $\lVert ( \mathcal H - E)\ket \psi\rVert \leq \Delta' < \Delta$. 
    Then: 
    \begin{align}
        \left| \bra{\Psi}  \psi\rangle \right|^2 \geq 1 - \frac{\Delta'^2}{\Delta^2}\label{eq:overlap_lower_bound}
    \end{align}
\end{lemma}
Assuming there exists $\ket \psi$ such that there exist $\alpha,$ $\beta$ such that inequality \eqref{kato_inequality_2} holds, then we can get an example of value for $\Delta'$ which depends on $\mathcal R$, and reads:
\begin{align} 
     \Delta'^2 \equiv \mathcal R^2 + \max \left( \frac{\mathcal R^2}{\mathcal E - \alpha}, \frac{\mathcal R^2}{\beta - \mathcal E} \right)^2 \label{eq:delta-prime-ref}
\end{align}
Indeed, we can see that:
\begin{align}
    \lVert (\mathcal H - E)\ket \psi\rVert^2 = \mathcal R^2 + (E-\mathcal E )^2
\end{align}
And thus we just need to bound $(E-\mathcal E )^2$ using Eq. \eqref{kato_inequality_2} to obtain Eq. \eqref{eq:delta-prime-ref}.

\subsubsection{Looking at two special cases} \label{bound_special_case}
Assuming that there is indeed only one eigenvalue between $ \mathcal E - \mathcal R$ and  $\mathcal E + \mathcal R$, and that this eigenvalue is separated from the rest of the spectrum by a distance $\Delta > 2 \mathcal R$, we can see that we get the bounds:
\begin{align}
    \left|  \mathcal E - E\right|&\leq \mathcal R \\
    \left| \bra{\Psi} \psi \rangle \right|^2& \geq 1 - \frac{2\mathcal R^2 }{\Delta^2} 
\end{align}
Where $\Delta$ is the gap between $E$ and the rest of the spectrum. This is already a rather interesting bound since it means $\mathcal E$ and $\ket \psi$ converge towards $E$ and $\ket \Psi$ as $\mathcal R$ approaches $0$. \\
In the case where Inequality \eqref{R-spectral-gap} holds, we get the bound:
\begin{align}
   \left| \mathcal E - E \right| &\leq \frac{2\mathcal R^2}{\Delta},  \\
   \left| \bra{\Psi} \psi\rangle \right|^2 &\geq 1 - \frac{\mathcal R^2 }{\Delta^2} - \frac{4 \mathcal R^4}{\Delta^4},
\end{align}
which this time involves $\mathcal R^2$ instead of $\mathcal R$.
Finally, assuming we are at the lower end of the spectrum, i.e $\mathcal E$ is smaller than the second lowest eigenvalue of $\mathcal H$ which we write $E_1$, we can set $\alpha = -\infty$ and $\beta = E_1$ and we get Temple's inequality which will be the focus of the next subsection.

\subsubsection{Case of the ground state: Temple's Inequality} \label{temple-sec}

In this subsection we introduce a special case of Kato's Inequality meant for the lowest eigenvalue, and apply it to the problem at hand.

\begin{theorem}{Temple's Inequality \cite[Theorem 1]{harrell1978generalizations}} \label{temple_inequality} \\
    Let $\mathcal{H}$ be a self-adjoint operator on a Hilbert space with sesquilinear inner  product $\braket{\cdot | \cdot}$. Suppose that $E_\mathrm{gs}$,  the lowest eigenvalue of $\mathcal{H}$,  is isolated from the rest of the spectrum $\mathrm{Sp}(\mathcal{H})$ and define:
    \begin{align}
        E_{\mathrm{es}} \equiv \inf\left(\mathrm{Sp}(\mathcal{H})\backslash \{E_{\mathrm{gs}}\} \right)
    \end{align}
    Let $\ket{\psi}$ be a normalized vector of our Hilbert space such that $\bra{\psi}\mathcal{H}\ket{\psi} < E_{\mathrm{es}}$. Then:
    \begin{align} \label{temple_inequality_2}
        E_{\mathrm{gs}} \geq \bra{\psi}\mathcal{H}\ket{\psi} -  \frac{\bra{\psi}\mathcal{H}^2\ket{\psi}  - \bra{\psi}\mathcal{H}\ket{\psi}^2}{E_{\mathrm{es}} - \bra{\psi}\mathcal{H}\ket{\psi}}
    \end{align}
    Note that the variationality of the algorithm also yields:
    \begin{align}
        E_{\mathrm{gs}} \leq \bra{\psi}\mathcal{H}\ket{\psi}
    \end{align}
    \end{theorem}

Namely, Temple's inequality gives us a lower bound on the energy consisting in the trial state energy minus the variance of the energy divided by an energy gap. 

=

\subsection{Application to the projective quantum eigensolver} \label{application_to_pqe}

In the case of PQE, we are interested in bounding the distance between the actual ground state energy $E_{\mathrm{gs}}$ and the energy $\bra{\Phi_0}\U^{\dag}(\bm{t}) \H \U(\bm{t}) \ket{\Phi_0}$ of our trial state. We are going to apply Eq. \eqref{temple_inequality_2} to PQE by replacing $\mathcal{H}$ by $\U \H \U^\dagger$ and $\ket\psi = \ket{\Phi_0}$ (usually the Hartree-Fock state). 

\subsubsection{PQE energy bounds and overlap lower bound}

With our definitions: 
\begin{align}
    \U^\dag(\bm{t}) \H \U(\bm{t}) \ket{\Phi_0} = E_0(\bm{t})\ket{\Phi_0} + \sum_{\mu} r_\mu(\bm{t})\ket{\Phi_\mu},
\end{align}
where the sum runs on the entire (exponential in size) orthonormal basis. 
We thus get the numerator:
\begin{align}
    \nonumber& \bra{\Phi_0}\left(\U^{\dag}(\bm{t}) \H \U(\bm{t})\right)^2 \ket{\Phi_0} - \\ &\qquad \bra{\Phi_0}\U^{\dag}(\bm{t}) \H \U(\bm{t}) \ket{\Phi_0}^2 = \sum_{\mu} r_\mu(\bm{t})^2,
\end{align}
namely the variance is given by the sum of the residues squared.

Assuming $E(\bm{t}) < E_{\mathrm{es}}$ (for this to hold, it is enough for our starting point to have a lower energy that the first excited energy) and applying Theorem \ref{temple_inequality}, we are left with the error bound:
\begin{align}\label{eq:energy_error}
    E(\bm{t}) - \frac{\sum_{\mu} r_\mu(\bm{t})^2}{E_{\mathrm{es}} - E(\bm{t})} \leq E_{\mathrm{gs}} \leq E(\bm{t}).
\end{align}
Next, we apply Eqs.~\eqref{eq:overlap_lower_bound}-\eqref{eq:delta-prime-ref} to get the following lower bound on the overlap:
\begin{align}\label{eq:overlap_bound_pqe}
    |\bra{\phi_0(\bm t)}&\Psi\rangle|^2 \geq  & \\
   &\nonumber  1 - \frac{\sum_\mu r_\mu(\bm t)^2}{\left(E_{\mathrm{es}} - E_{\mathrm{gs}}\right)^2}
   \left( 1 +  \frac{\sum_{\mu} r_\mu(\bm{t})^2}{\left(E_{\mathrm{es}} - E(\bm{t})\right)^2} \right)
\end{align}
We will from now on use the notation:
\begin{align} 
    \varepsilon_\mathrm{T}(t) = \frac{\sum_{\mu} r_\mu(\bm{t})^2}{E_{\mathrm{es}} - E(\bm{t})}.\label{eps_t}
\end{align}

\subsubsection{Practical use of these bounds}
The energy and overlap bounds Eqs.~\eqref{eq:energy_error}-\eqref{eq:overlap_bound_pqe} have a number of interesting implications. 

The first one is that the smaller the squared residues ($\sum_{\mu} r_\mu(\bm{t})^2$), the closer we are to the ground-state energy and the larger the overlap with the ground state.
Reducing the residues therefore makes sense in order to get a better approximation to the ground state energy.
We can however expect that since there is an exponential (in the number of orbitals) amount of residues, for a given ansatz with a polynomial number of parameters, the quantity $\sum_{\mu} r_\mu(\bm{t})^2$ will be bounded by below, and thus so will be the best attainable accuracy.

Still, focusing on a polynomially sized set of terms $\mathcal{A}$ is a suitable approach, provided that $\sum_{\mu \notin \mathcal{A}} r_\mu(\bm{t^{(0)}})^2$ is small enough to allow the desired accuracy to be reached, and provided that this quantity will not grow too much during the optimization. There is a priori no reason the second point should hold, but the reasons why it is to be expected will become clearer after section \ref{math_section}.

The second interesting implication of Eq. \eqref{eq:energy_error} is practical. During the optimization, we have access to both $\sum_{\mu \in \mathcal{A}} r_\mu(\bm{t})^2$ and $E_0(\bm{t})$. If we suppose that our initial state is not too bad of an approximation of the ground state, we can suppose that the initial energy $E_0(\bm t^{(0)})$ ({\it e.g} the Hartree Fock energy) is such that $E_{0}(\bm t^{(0)}) < E_\mathrm{es}$ or at least $E_{0}(\bm t^{(0)}) \approx E_\mathrm{es}$. To shorten notation, we will from now on write $E_0 \equiv E_0(\bm t^{(0)})$. This leads to the approximate bound:
\begin{align} \label{hf_error_bound}
   \varepsilon_{\mathrm{exact}}(t) \equiv |E(\bm{t}) - E_{\mathrm{gs}} | \leq \frac{\sum_\mu r_\mu(\bm{t})^2}{E_0 - E(\bm{t})}. 
\end{align}

We are still missing the contribution from residues outside our operator set $\mathcal{A}$ ($\sum_{\mu \notin \mathcal{A}} r_\mu(\bm{t})^2$) in order to assess the error.
Nevertheless, as stated earlier, one can only hope that the set $\mathcal{A}$ is good enough, such that at the start this term will be negligible with respect to the target accuracy, and that it will remain so during the optimization.
Note that Ref.~\onlinecite{Stair_2021} presents a way to choose $\mathcal{A}$ based on the residues, and that using this method makes it theoretically possible to have an estimation of $\sum_{\mu} r_\mu(\bm{t})^2$. 

Under those assumptions, we can derive a rather well informed convergence criterion of the form:
\begin{align} \label{convergence_criterion}
    \varepsilon_\mathrm{T}^{\mathcal{A}}(t)\equiv \frac{\sum_{\mu \in \mathcal{A}} r_\mu(\bm{t})^2}{E_{0} - E_0(\bm{t})} < \varepsilon
\end{align}
where $\varepsilon$ is a user-defined threshold. 

Lastly, we would like to point out that the numerator in the rhs. of Eq. \eqref{temple_inequality_2} corresponds to the variance of the energy.
Therefore, one can see the similarity of PQE and variance-based methods such as Ref.~\onlinecite{zhang2020variationalquantumeigensolversvariance}: there, convergence towards an eigenstate is guaranteed by minimizing the variance of the energy.
In PQE, we are also interested in minimizing the variance, but only look at a portion of it.
This has two upsides. The first is that looking at a carefully chosen portion of it is less costly than computing it entirely. The second one is that knowing components of the variance allows us to use them in order to reduce it, meaning we can perform advanced optimization techniques (such as the one introduced in section \ref{sec:Better_opti}), the fast convergence of which will also help bring down the overall cost of the algorithm.

\subsubsection{Sensitivity to barren plateaus}

Barren plateaus refer to regions on a manifold where the variance and gradients of the loss function vanish exponentially with the size of the system. In our case, though we do not use gradients, we still subtract energies in order to compute residues, and thus if the landscape is almost flat, the cost in shots in order to have sufficient precision on the energies can become prohibitive. However, several elements suggest that this might not be as severe a problem in our case. 

First, having a convergence criterion linked (albeit approximately) to the error, contrary to VQE, means we know in advance the accuracy we need on the residues.
If the optimization terminates, we can assume with good confidence that we are indeed close to an eigenvalue (which is not the case with gradients).
Moreover, the energies we measure are technically values at one point of $3$ different manifolds (although generated by the same unitary), and since we initialize our ansatz at $\bm{t}=0$, around which the landscape does not exhibit barren plateaus \cite{mhiri2025unifyingaccountwarmstart}, there is a good chance we will not face such problems.
(A caveat is that, in systems with strong correlations, the ground state will likely be away from $\bm{t}=0$ and thus the optimization will have to travel from one fertile valley to another.)
If the rough statement $\lVert r(t) \rVert \approx 0 \Rightarrow E(t) - E_{\mathrm{gs}}\approx 0$  still holds (provided $\lVert r(t) \rVert$ is smaller than the energy gap), we might run into problems different from barren plateaus. For instance, it might be hard for the optimizer to reach convergence. Such cases arise even in low dimensions, and are discussed in section \ref{math_section}.

\subsection{Numerical checks}
After this analytical derivation of the bounds, we check them numerically on a few example systems. One aim is to test their tightness.

\subsubsection{Complete bound}

 \begin{figure}
     \centering
     \includegraphics[width=0.5 \textwidth]{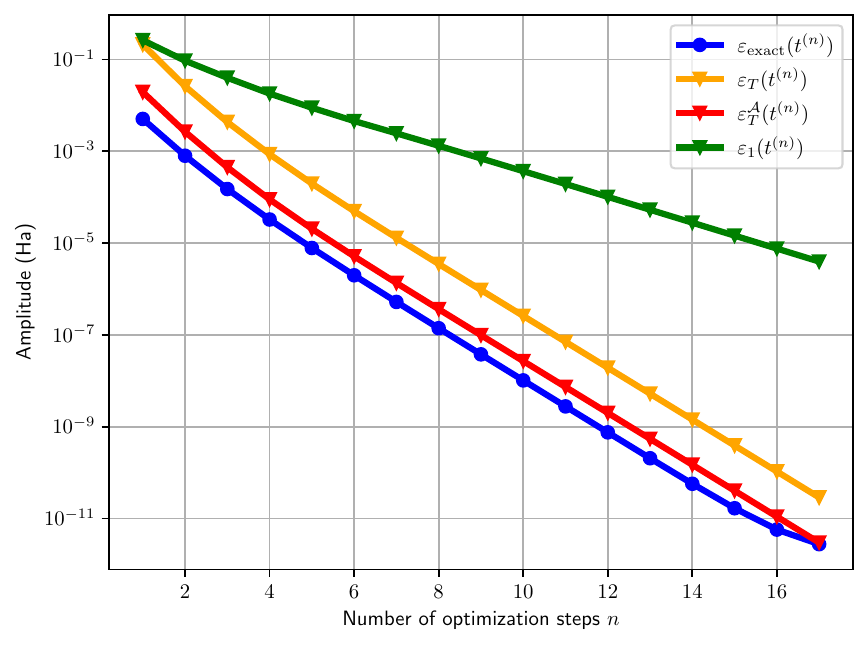}
     \caption{Comparison of the exact error $\varepsilon_\mathrm{exact}$, our bound $\varepsilon_{\mathrm{T}}$, our more easily computable bound $\varepsilon_{\mathrm{T}}^{\mathcal A}$ and the bound from \cite{Stair_2021}, by running the PQE algorithm on an $\mathrm{H_4}$ molecule at $d_{H-H} = 0.75\text{\AA}$ using a complete UCCSDTQ ansatz. The threshold for convergence $\varepsilon$ (Eq. \eqref{convergence_criterion}) was set to $10^{-10}$.}
     \label{error_bound_check}
 \end{figure}

In order to test the complete bound (Eq.~\eqref{eq:energy_error}), we ran the PQE algorithm on a $\mathrm{H}_4$ molecule at $d_{H-H} = 0.75\text{\AA}$ using a complete UCCSDTQ ansatz. The initial state was chosen as the Hartree Fock state, which corresponds to setting $\bm t^{(0)} = 0$. Simulations were run using our own implementation of the algorithm on Eviden's Qaptiva platform, and in particular the myQLM-fermion library \cite{Haidar2022}.
One and two-body integrals were computed using PySCF using a sto-$3$g basis. The exact values of $E_{\mathrm{gs}}$ and $E_{\mathrm{es}}$ where obtained via exact diagonalization (full configuration interaction). 

The results are shown in Fig.~\ref{error_bound_check}.
The blue slope  represents the exact error $\varepsilon_\mathrm{exact}$ (defined in Eq.~\eqref{hf_error_bound}).
The red slope  represents the exact error upper bound (based the exponentially large set of all operators), $\varepsilon_{\mathrm{T}}(\tn)$.
The orange slope represents the estimate of the error upper bound based on the operators of the (in this case complete) polynomial set $\mathcal{A}$, $\varepsilon_{\mathrm{T}}^\mathcal{A}(\tn)$, which is the only estimate we can compute in practice.
Finally, the green slope represents the bound mentioned in Ref.~\onlinecite{Stair_2021}, which bounds the energy error with the residual $1$-norm $\varepsilon_1(\bm{t}) \equiv \sum_{\mu} |r_\mu(\bm{t})|$.

We observe that
our exact bound $\varepsilon_{\mathrm{T}}(\tn)$ (provided by Eq. \eqref{hf_error_bound}) provides a convincing estimation of the energy error as it is off by less than $2$ orders of magnitude from $\varepsilon_\mathrm{exact}$, to be compared to the residual $1$-norm, which is off by between $2$ and $6$ orders of magnitude.
Our approximate bound $\varepsilon_{\mathrm{T}}^\mathcal{A}(\tn)$ is about half an order of magnitude larger than the true value of the energy error.

\subsubsection{Convergence criterion} \label{criterion_benchmarks}

In this section, we use the approximate upper bound $\varepsilon_{\mathrm{T}}^\mathcal{A}(\tn)$ as a stopping criterion (see Eq. \eqref{convergence_criterion} above) for the PQE algorithm:
we ran the PQE algorithm as described in section \ref{brief_summary}, using different values for $\varepsilon$ and at different bond distances $d_{H-H}$, for the LiH and H$_4$ molecule.
 
\begin{figure}[H]
    \centering
    \includegraphics[width=0.5 \textwidth]{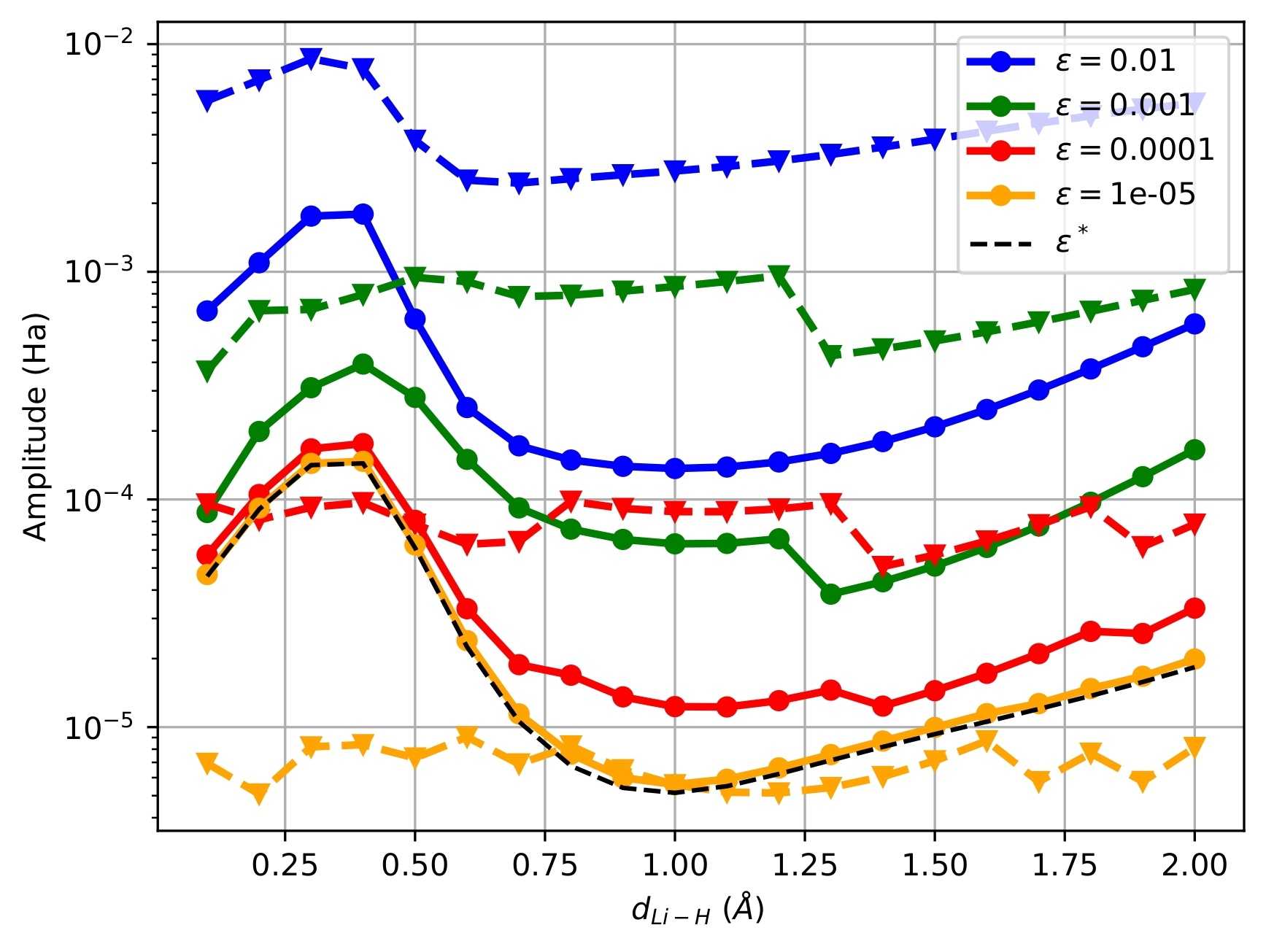}
    \caption{Test of our convergence criterion on LiH: $\varepsilon_\mathrm{T}$ (solid lines) and $\varepsilon_\mathrm{T}^\mathcal{A}$ (dashed lines) as a function bond distance $d_{H-H}$ for different values of the threshold $\varepsilon$, with a UCCSD ansatz.}
    \label{criterion_check_2_simple}
\end{figure}

\begin{figure}[H]
    \centering
    \includegraphics[width=0.5 \textwidth]{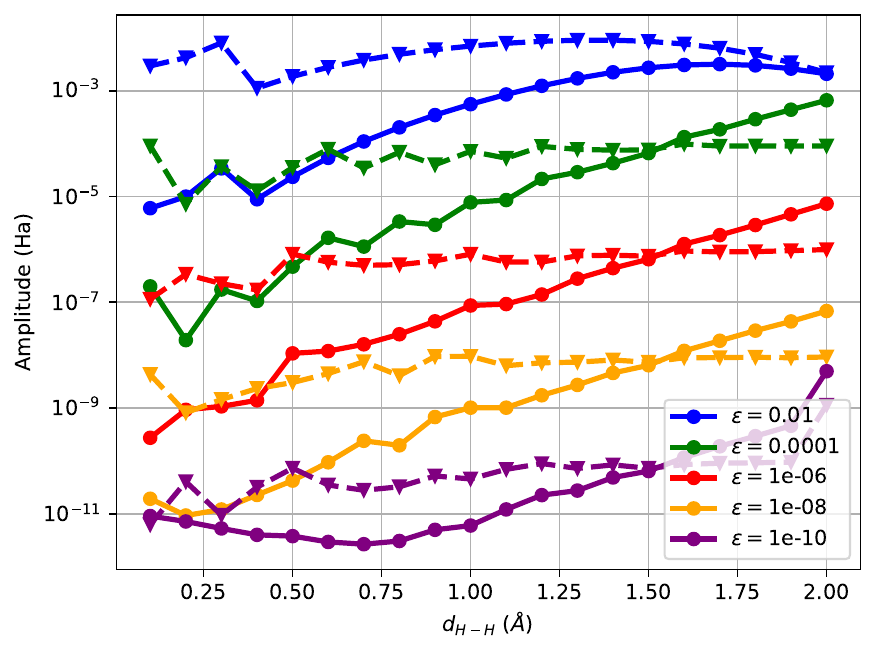}
    \caption{Same as Fig.~\ref{criterion_check_2_simple} for $\mathrm{H_4}$, for a UCCSDTQ ansatz}
    \label{criterion_check_3_simple}
\end{figure}

 We introduce the notations: 
 \begin{align}
 \varepsilon_{\mathrm{exact}}^* &= E_0(\bm t_\mathrm{final}) - E_{\mathrm{gs}}, \\
     \varepsilon^* &= \min_{\bm t} E_0(\bm t) - E_{\mathrm{gs}},
 \end{align}
which correspond respectively to the error obtained at the end of the algorithm (and which depends on the threshold $\varepsilon$) and to the lowest energy error attainable within the manifold.
 The results are shown in Figs. \ref{criterion_check_2_simple} and \ref{criterion_check_3_simple}.
 
 For each value of the bond distance, and $\varepsilon$, we plot the final value of $\varepsilon_\mathrm{T}^\mathcal{A}$ with a dashed line, and $\varepsilon_\mathrm{exact}^*$. The black dashed line represents $\varepsilon^*$, which was calculated by running VQE with the exact same UCCSD ansatz (in which second-order operators are put to the right of first-order ones).


We can see from Figure \ref{criterion_check_2_simple} that our criterion does a good job at guaranteeing a given accuracy $\varepsilon$ when $\varepsilon^* \ll \varepsilon $.
By looking at Eq. \eqref{eq:energy_error}, one can see that the lower bound $\varepsilon^*$ is coming from the residues  which are not optimised during PQE ($r_\mu$ with $\mu\notin \mathcal{A}$). 
We would however like to point out that in highly correlated cases, there can be a slight discrepancy between the best attainable error $\varepsilon^*$ and the error yielded by PQE with a threshold $\varepsilon \ll \varepsilon^*$. This discrepancy is explained by the fact that residues corresponding from operators not in the ansatz are slightly influenced by the choice of parameters, and thus there can be a place $t^*$ in the manifold, where $\bm t^* =  \argmin_{\bm t}  E_0(\bm t) - E_{\mathrm{gs}} $, such that $\varepsilon^\mathcal{A}_\mathrm{T}(t^*) \neq 0 $. 

When looking at Figure \ref{criterion_check_3_simple}, we can see that up to a value $d_{H-H} \approx 1.5 \text{\AA}$, our criterion does a good job at guaranteeing a given accuracy $\epsilon$ (which can always be reached since we are using a complete ansatz). 
For low bond distances the discrepancy between $\varepsilon_\mathrm{exact}$ and $\varepsilon_\mathrm{T}^\mathcal{A}$  is rather large, while for bond distances $> 1.5 \text{\AA}$, we no longer have $\varepsilon_\mathrm{exact}<\varepsilon_\mathrm{T}^\mathcal{A} $. This comes from the fact that at low bond distances, there is a large gap between $E_{\mathrm{es}}$ and the Hartree Fock energy $E_0$ (meaning the bound is looser), and at large bond distances $E_{\mathrm{es}} < E_0$.

\section{Mathematical focus on quasi-Newton methods for PQE} \label{math_section}

Contrary to VQE where we are interested in minimizing the trial state energy, in PQE we want to find the zeros of the residue map. Common methods for approximately solving this problem are quasi-Newton methods, which will be the focus of this section. Our purpose is to introduce tools which we will later use in order to improve,  both in terms of robustness and convergence speed, the optimization method proposed in Ref.~\onlinecite{Stair_2021}.

\subsection{On quasi-Newton methods}
\subsubsection{General discussion on update rules} \label{discussion_update_rules}
The goal of this section is to provide the intuition leading to what will later on be properly mathematically introduced. 

Given a map $F: \bm{t} \in \mathbb{R}^N \rightarrow F(\bm{t})$, the question of finding $\bm{t}^*$ such that $F(\bm{t}^*) = 0$ is a highly nontrivial one. A common approach to such problems is the Newton-Raphson algorithm, which aims at iteratively approximately solving the root equation. Provided $F$ is differentiable with Jacobian map $J$, the algorithm starts with an initial guess $\bm{t}^{(0)}$  and updates $\bm{t}^{(n)}$ following the update rule:
\begin{align} \label{newton-raphson}
    \tnn = \tn - J(\tn)^{-1} F(\tn)
\end{align}
This requires each $J(\tn)$ to be non-singular. 

The motivation for such an update rule is the following: suppose that for a given $t$ one can linearly approximate  $F(\bm{t} + \Delta \bm{t}) - F(\bm{t})$, meaning one can write $F(\bm{t} + \Delta \bm{t}) - F(\bm{t}) \approx A(\bm{t})\Delta \bm{t}$. Then if 
$A(\bm{t})$ is non-singular we get for $\Delta \bm{t} = -A(\bm{t})^{-1}F(\bm{t})$: 
\begin{align}
    F(\bm{t}+ \Delta \bm{t}) \approx F(\bm{t}) -  A(\bm{t})(A(\bm{t})^{-1}F(\bm{t})) = 0
\end{align}
Therefore, if the linear approximation is good, we should expect $F(\bm t+ \Delta \bm t)$ to be small. For $F(\bm t+ \Delta \bm t)$ to be the smallest possible, it makes sense to use the best linear approximation of $F(\bm t + \Delta \bm t) - F(\bm{t})$, which for a small $\Delta \bm t$ is known to be obtained by choosing $A(\bm{t}) = J(\bm{t})$. 

The quality of the update rule given by \eqref{newton-raphson} is highly dependent on how well $\Delta t \rightarrow J(\bm{t}) \Delta t$ approximates $F(\bm t + \Delta \bm t) - F(\bm{t})$. 

Now, without any straightforward way to compute $J(\bm{t})$, or if the cost of inverting it is too high, one can use an approximation $A$ of it. Then, provided that this approximation is good enough (ie close to $J(\bm{t})$), then the update rule:
    \begin{align}
    \tnn = \tn - A(\tn)^{-1} F(\tn)
    \end{align}
should still perform well (provided Newton-Raphson does). 

This is the idea behind the \textit{modified} Newton-Raphson algorithm, which replaces $J(\bm{t})^{-1}$ with $J(\bm{t}^{(0)})^{-1}$ to avoid the computational cost of computing and inverting $J(\bm{t})$ at each step. The idea is simply that if the optimization does not go too far from $\bm t^{(0)}$ (meaning the solution is close to the initial guess), $J(\bm{t}^{(0)})$ should remain somewhat close to $J(\bm{t})$. This yields the update rule:
\begin{align}
    \tnn = \tn - J(\bm{t}^{(0)})^{-1}F(\tn).
\end{align} 

One of the caveats of the Newton-Raphson algorithm is that the Jacobian matrix $J(\bm{t})$ is only the best linear approximation of the variations of $F$ \textit{around} $\bm t$, meaning that for the algorithm to converge, an optimization step $\Delta \bm t$ should be small enough for $J(\bm{t})$ to approximate $F(\bm t + \bm \Delta t) - F(\bm{t})$ well. Therefore, one might encounter cases where $\Delta \bm t$ is too big and thus $\Delta \bm t \rightarrow F(\bm t + \Delta \bm t) - F(\bm{t})$ is very poorly approximated by $\Delta \bm t \rightarrow J(\bm{t})\Delta \bm t$. A naive workaround is to add a variable step size chosen via a line search on the norm of $F$. This can however lead to slow convergence, and adding a line search can make each step much more computationally expensive. In such cases, it can also be better to use a linear approximation of $\Delta t \rightarrow F(\bm t + \Delta \bm t) - F(\bm{t})$ which is worse around $\bm t$, but will overall yield better approximations of $\Delta \bm t \rightarrow F(\bm t + \Delta \bm t) - F(\bm{t})$. 
We will see in the next sections that such cases do exist in PQE and that there are workarounds in order to optimize the residues. 
Note that in what precedes we hide the subtlety induced by the fact that the size of the steps we will perform will depend heavily on the approximation we use. 

The next section will introduce the Newton-Kantorovich theorem, which provides a convergence criterion for the Newton-Raphson and modified-Newton-Raphson algorithms, allowing us to have a proper mathematical framework to complete our intuition.
\subsubsection{The Newton-Kantorovich theorem}
First stated by Kantorovich in 1948, the Newton-Kantorovich theorem has since been refined and generalized. We here introduce the version proven in Ref.~\onlinecite{Argyros_2007}:
\begin{theorem}[Newton-Kantorovich Theorem] \label{newton-kantorovich-theorem}
    Let $t_0 \in \mathbb{R}^N$, let $F:\mathcal{B}\left(\bm t_0, R\right) \rightarrow \mathbb{R}^N$ a Frechet-differentiable operator, let $\bm t \rightarrow J(\bm{t})$ its Jacobian map. \\
    Suppose:
    \begin{align} \label{Kantorovich-nonsingular-jacobian}
        J(\bm t_0) \in \mathrm{GL}_N(\mathbb{R})
    \end{align}
    and there exists $\eta, \  L > 0$ such that:
    \begin{align} \label{kantorovich-eta-L-def}
    &J(\bm t_0)^{-1} F(\bm t_0) \leq \eta, \\ \label{kantorovich_first_step}
    \text{For all } \bm t \in \mathcal{B}\left(\bm t_0, R \right), \\ \nonumber \ \lVert &J(\bm t_0)^{(-1)}J(\bm{t}) - I \rVert \leq L \lVert \bm t- \bm t_0\rVert, \\
    &\eta L < \frac12, \label{kantorovich_convergence_criterion} \\ \label{kantorovich-lipschitz}
    &\frac{1 - \sqrt{1-2L\eta}}{L} \leq R.
    \end{align}
    Then, the \textit{modified Newton-Raphson} sequence defined by:
    \begin{align} \label{modified-newton-update-rule}
    \bm t^{(0)} &= \bm t_0 \\
    \tnn &= \tn - J(\bm t_0)^{-1}F(\tn)
    \end{align}
    converges to a unique root solution $t^*$ of the equation $F(\bm{t}) = 0$.
    Moreover, we have the following estimates for $n \geq 0$:
    \begin{align} \label{convergence_estimates}
        &\lVert \tnn - \tn \rVert \leq q^n \lVert \bm t^{(1)} - \bm t^{(0)} \rVert, \\
        &\lVert \tn - \bm t^* \rVert \leq \frac{q^n}{1-q} \eta = \frac{q^n}{1-q} \lVert \bm t^{(1)} - \bm t^{(0)} \rVert ,
        \end{align}
        where $q = 1 - \sqrt{1-2\eta L}$ \\ \\
    Additionaly, if we have:
    \begin{align}
        \text{For all } \bm t \in \mathcal{B}\left(\bm t_0, R\right),\  J(\bm{t})  \in \mathrm{GL}_N(\mathbb{R})
    \end{align}
    Then, the \textit{Newton-Raphson} sequence defined by:
    \begin{align} \label{newton-update-rule}
    \mathbf{\bm t}^{(0)} &= \bm t_0 \\
    \mathbf{\bm t^{(n+1)}} &= \mathbf{t^{(n)}}- J(\mathbf{\bm t^{(n)}})^{-1}F(\mathbf{t^{(n)}})
    \end{align}
    converges to a unique root solution $t^*$ of the equation $F(\bm{t}) = 0$.
\end{theorem}
The Newton-Kantorovich Theorem provides both qualitative and quantitative insights into the Newton-Raphson and modified Newton-Raphson algorithms. 
Moreover, the hypotheses of the theorem properly formulate the intuitions (such as the ones we discuss in \ref{discussion_update_rules}) one can have on quasi-Newton methods.

Theorem \ref{newton-kantorovich-theorem} also contains statements about convergence speed which will be the focus of subsection \ref{subsec:convergence}.

While \eqref{kantorovich_first_step} encompasses the size of the first optimization step, which can be seen as an indicator of how far from the initial guess the optimization might go wrt. how much $F$ varies, Eq. \eqref{kantorovich-lipschitz} can be seen as a weaker Lipschitz-continuity condition for $\bm t \rightarrow J(\bm t_0)^{-1}J(\bm{t})$. Intuitively, since one wants $F$ to be as 'linear' as possible, in which case its Jacobian map would be constant, and the constant $L$ (which is essentially a 'weaker' Lipschitz constant) can be seen as an indicator of how  constant the Jacobian map is, and therefore as an indicator of how far $F$ is from a linear map. Mathematically, $L$ plays a role in how contracting the update map is (the Lipschitz constant of the update map being $q$, see the proof of Theorem 5 in Ref.~\onlinecite{Argyros_2007}). The lower $L$, the more contracting it is. The update map being contracting is what provides the solution existence and uniqueness, and the more contracting it is, the faster the convergence of the algorithm. 

Noticing that the optimization algorithm for the PQE introduced in Ref.~\onlinecite{Stair_2021} can be seen as an approximation of a modified Newton-Raphson algorithm, one can hope to use this theorem to obtain insights into the convergence properties of the algorithm. 

The next sections focus on introducing several update rules for PQE and on studying their convergence properties using the tools we have introduced.

\subsection{Proof of concept: deriving a hybrid update rule}

In our previous section, we introduced general formulas for quasi-Newton update rules, mainly the well known Newton-Raphson update rule (Eq. \eqref{newton_update_rule}, and the lesser known modified-Newton-Raphson (Eq. \eqref{modified-newton-update-rule}). Both update rules require knowledge on the Jacobian map of our operator, the former requiring the computation of the Jacobian matrix at each iteration while the latter only requires the Jacobian matrix at the initial point. More generally, quasi-Newton update rules require some linear approximation of the variations of our operator. 

In the case of the residue map, which is the operator whose roots we want to find, computing the Jacobian matrix is however highly non-trivial. We introduce in this section two update rules aiming at approximating a Newton-Raphson and a modified-Newton-Raphson optimization. We then show that using Theorem \ref{newton-kantorovich-theorem}, we can combine it with the update rule from Ref.~\onlinecite{Stair_2021} to obtain better convergence properties. 

\subsubsection{The update rules}
The update rule advocated by Stair \& Al. Ref.~\onlinecite{Stair_2021}, Eq.~\eqref{standard-update-rule}, the derivation of which can be found in \ref{jacobian-derivations}, can be seen as an approximation of a modified Newton-Raphson update rule with $\bm t^{(0)} = 0$ (the starting point is the classically obtained Hartree Fock state) and with $J(0) \approx \delta_{\mu,\nu} \Delta_\mu$.

This latter approximation amounts to neglecting correlation terms in the Hamiltonian. One can thus see that it is possible to obtain a systematically better approximation of $J(0)$ by taking into account correlation terms affecting the diagonal of the Hamiltonian (in other terms, by approximating $\H$ with its entire diagonal instead of taking only a part of it).
This leads to the quite similar update rule:
\begin{align} \label{approximate_modified}
     \tnn_\mu = \tn_\mu + \frac{r_\mu(\tn)}{E_0(0) - E_\mu(0)}.
\end{align}
In cases where the solution is fairly close to our starting point, we have found in our simulations that using this update rule yielded a better convergence rate than the rule from Eq. \eqref{standard-update-rule}. However, this approximation results in the idea that getting a better approximation of $J(0)$ will yield a better linear approximation of the residue map during the optimization.
This is however not the case when the solution lies in a region where $J(0)$ is a bad linear approximation. Thus, this approximation makes sense only when $J(0)$ is overall a good linear approximation of the residue map in the region in which we are interested.
For instance, in cases where correlation is strong, the solution to the residual equations lies far from the Hartree Fock state, and thus a linear approximation based on the Jacobian matrix at $0$ is not good enough for convergence to take place.
Intriguingly, we have found empirically that neglecting the correlation terms  altogether (namely using $ \diag\left(-\delta_{\mu,\nu} \Delta_\mu \right)_{\mu, \nu\in \mathcal{A}}$ in place of $J(0)$, see \ref{math_section} for more details) leads to an overall decent linear approximation of the variations in some of those cases.

 If we now want to approximate a Newton-Raphson update rule (as opposed to modified Newton-Raphson), one can show (see Appendix \ref{jacobian-derivations}) that $J(\bm{t}) \approx \left(E_\mu(\bm{t}) - E_0(\bm{t})\right)_{\mu, \nu\in  \mathcal{A}}$, leading to the update rule:
 \begin{align} \label{approximate_newton}
     \tnn_\mu = \tn_\mu + \frac{r_\mu(\tn)}{E_0(\tn) - E_\mu(\tn)}.
\end{align}
Since this rule approximates $J(\bm{t})$ instead of $J(0)$, one can expect it to converge faster than \eqref{approximate_modified}. However, this update rule is still based on the assumption that the solution will not be too far from the starting point, and thus the Jacobian map will yield a good linear approximation of the variations of the residue map close to the solution, which is not always the region of interest. We will show in section \ref{benchmarks_hybrid} that the update rule from Eq. \eqref{standard-update-rule} yields better stability than this one. 
  
Following the convergence properties of both the update rules from Eq. \eqref{standard-update-rule} and Eq. \eqref{approximate_newton}, it is only natural to want to combine the good properties of both, mainly:
\begin{itemize}
    \item The higher convergence speed of the approximate Newton-Raphson update rule
    \item The better stability of the standard update rule.
\end{itemize}
The next subsection is devoted to using Theorem \eqref{newton-kantorovich-theorem} in order to obtain an update rule combining both.

\subsubsection{Deriving the hybrid update rule}
We discussed in section \ref{discussion_update_rules} three update rules, the two of interest being the approximate Newton-Raphson update rule (Eq. \eqref{approximate_newton}), and the standard update rule (Eq. \eqref{standard-update-rule}), the pros and cons of which can be summarized as:
\begin{itemize}
    \item An approximation of a Newton-Raphson (\ref{approximate_newton})  optimization is faster than the standard update rule \eqref{standard-update-rule} when it converges.
    \item The standard update rule converges in more cases.
\end{itemize}
One can show (see Appendix \ref{jacobian-derivations}) that we have the approximate inequality for $\bm t_1, \ \bm t_2 \ll 1$:
\begin{align}
    \lVert J(\bm t_1)^{-1} (J(\bm{t}_2) - J(\bm t_1) \rVert_1 \leq \lVert \bm{t}_1 - \bm t_2 \rVert_1 .
\end{align}
This means that the constant $L$ is approximately equal to $1$. In turn, this implies the convergence condition \eqref{kantorovich_convergence_criterion} from Theorem \ref{newton-kantorovich-theorem} now reads:
\begin{align}
    &\eta < \frac12 \\
    \text{i.e }& \lVert J(\bm t^{(0)})^{-1}r(\bm t^{(0)}) \rVert_1 < \frac12 
\end{align}
Therefore, we can approximately predict when a Newton-Raphson optimization will converge by looking at the $1$-norm of the first optimization step. 
In the context of the approximate Newton-Raphson update rule, this reads:
\begin{align}
    \sum_{\mu\in \mathcal{A}} \left| \frac{r_\mu(\bm t^{(0)})}{E_{\mu}(\bm t^{(0)}) - E_0(\bm t^{(0)})} \right| < \frac12
\end{align}
It thus seems natural to use this criterion to select the best update rule at each step: we start the optimization using $\eqref{standard-update-rule}$ until the convergence criterion is met (ie we are close enough to the solution), and then to take advantage of the faster convergence of \eqref{approximate_newton}. 

This leads to the update rule which we dub 'hybrid': 
\begin{align} \label{hybrid_update_rule}
     \tnn_\mu = \begin{cases} &\tn_\mu + \frac{r_\mu(\tn)}{\Delta_\mu} \  \text{if} \  \eta_n  > \frac12 \\
     &\tn_\mu + \frac{r_\mu(\tn)}{E_0(\tn) -E_\mu(\tn)} \ \text{otherwise}
     \end{cases}
 \end{align}
where: 
 \begin{align}
     \eta_n \equiv \Bigg\lVert\left(\frac{r_\mu(\tn)}{E_\mu(\tn) -E_0(\tn)}\right)_{\mu \in \mathcal{A}}\Bigg\rVert_1.
     \label{eq:eta_def}
 \end{align}
We will show in the next section that this update rule allows a faster convergence of the residues and is slightly more robust than the standard update rule \eqref{standard-update-rule}.

 \subsubsection{Benchmarks and discussion} \label{benchmarks_hybrid}
To compare the different optimization methods, we ran the algorithm using each method on Eviden's Qaptiva simulation platform, on $\mathrm{H_4}$, $\mathrm{H_6}$, and $\mathrm{BeH_2}$ molecules corresponding to $8$, $12$ and $14$ qubits, respectively, for bond distances ranging from $0.1 \text{\AA}$ to $4 \text{\AA}$. Hartree Fock calculations and one and two-body integral calculations where performed using PySCF with a sto-$3$g basis. The translation from fermionic to spin operators was performed using the Jordan-Wigner encoding \cite{Jordan1928}.
We use the exact same UCCSD ansatz for each optimization method, in which the ordering reads $\U(\bm t) = \U_1(\bm t) \U_2(\bm t)$ where $\U_1$ contains the first order excitation operators, and $\U_2$ contains the second order excitation operators. We use a convergence threshold $\varepsilon = 10^{-5}$ (as defined in Eq. \eqref{convergence_criterion}), and as usual a starting point $\bm{t}^{(0)} = 0$. Since we are interested in comparing the update rules themselves and their convergence, we did not use Direct Inversion in the Iterative Subspace\cite{PULAY1980393} (DIIS) here. It could be added in the same way to each optimization, and we found in our simulations that it did not affect much the optimization's ability to reach convergence (see Fig. \ref{LIH_vqepqe_diis} for a plot of the standard optimization with DIIS).

 We introduce two notations for simplicity: $\varepsilon_{\mathrm{final}}$ refers to the energy after convergence, while $\varepsilon^*$ refers to the lowest energy encountered during the algorithm.
 
\begin{figure*}[t!]
     \centering
     \includegraphics[width = 1.0\textwidth]{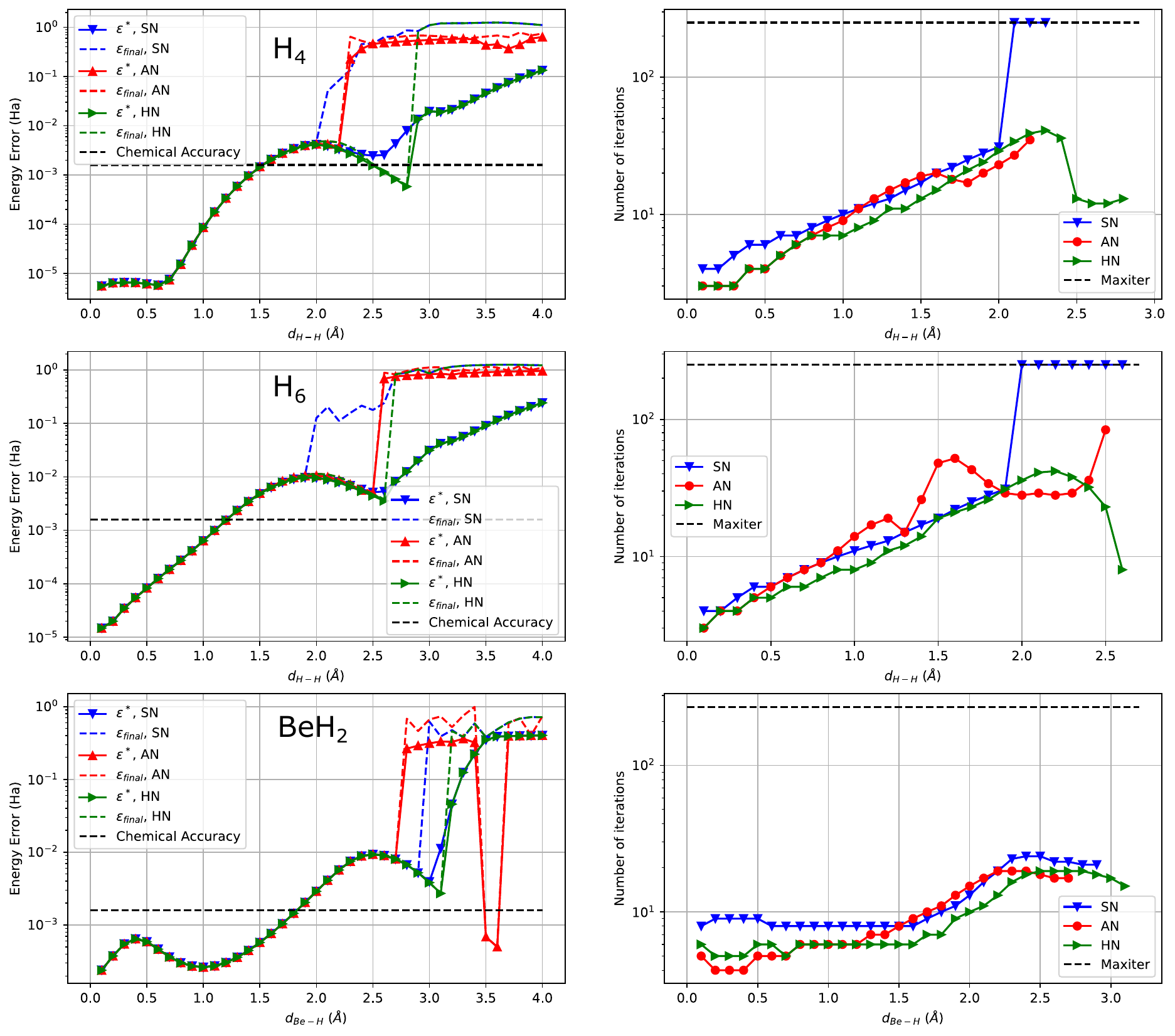}
     \caption{Comparison of the 'hybrid' quasi-Newton optimizer, the approximate Newton-Raphson optimizer, and the standard quasi-Newton optimizer for PQE ran on three molecules at different bond distances: $\mathrm{H}_4$ (top), $\mathrm{H}_6$ (middle) and $\mathrm{BeH}_2$ (bottom). Left panels: energy error, ie the difference between the energy value yielded by the algorithm and the exact ground state energy computed via FCI. Right panels: total number of optimization steps required to reach convergence. The $x$ axis of each plot represents the bond distance.}
     \label{H4_sn_hn}
 \end{figure*}

 When looking at Figure \ref{H4_sn_hn}, we can see that the hybrid update rule performs better on $\mathrm{H_4}$ than the standard update rule: it requires systematically fewer steps for convergence, and while the standard update rule stops converging around $2 \text{\AA}$, the hybrid update rule converges until $2.8 \text{\AA}$. As expected, the approximate Newton-Raphson update rule performs similarly to the hybrid update rule at low bond distances, but becomes worse between $1\text{\AA}$ and $1.6 \text{\AA}$, and even though it is then slightly better then, it stops converging sooner, around $2.2\text{\AA}$. This suggest that perhaps the threshold for the switch between the standard update rule and the approximate Newton-Raphson update rule could be increased, meaning the approximation $L \approx 1$ might be a bit of an overestimation. 
 
 On $\mathrm{H_6}$ (Figure \ref{H4_sn_hn}), we can see the same behavior: the hybrid update rule converges slightly faster than the standard one, and converges in more cases. The approximate Newton-Raphson update rule performs mostly worse than the hybrid update rule, and in some cases worse than the standard update rule. 
 
 Finally, on $\mathrm{BeH_2}$ (Figure \ref{H4_sn_hn}) we get the same behavior: the hybrid update rule is superior to the standard update rule. Again, the approximate Newton-Raphson update rule is mostly inferior to the hybrid update rule, but performs better in some cases, suggesting again that the value taken for $L$ could be lower (ie we could switch to the approximate Newton-Raphson update rule sooner). Note that there are two points where the approximate Newton-Raphson update rule converges but not the other ones. This can be explained by the fact that the initial point is slightly good enough  for a Newton-Raphson to converge, but $\eta > \frac12$ (again, this suggests the threshold $\frac12$ might be too conservative) and the Hartree Fock Hamiltonian approximates very poorly the Hamiltonian, and thus the standard update rule performs poorly meaning the first steps of the hybrid optimization will not bring us closer to the solution and thus the approximate Newton-Raphson part of the optimization will never start.
 
\subsection{Proof of concept $2$: convergence rate predictions} \label{subsec:convergence}
\subsubsection{The Theory}
Here, we wish to use the Newton Kantorovich Theorem in order to investigate the convergence of a modified Newton-Raphson optimization.
We have previously introduced the update rule corresponding to an approximation of a modified Newton-Raphson optimization (see Eq. \eqref{approximate_modified}), 
and as stated previously, using derivations found in Appendix~\ref{jacobian-derivations}, we show that in first order in $\bm t$ and $\bm t^{(0)}$, one gets:
\begin{align}
    &\lVert J(\bm t^{(0)})^{-1}J(\bm{t}) - I \rVert_1 \leq \lVert \bm t^{(0)} - \bm t \rVert_1
\end{align}
where for conciseness, $\lVert \cdot \rVert_1$ is either the $1$-norm on $\mathbb{R}^N$ or the associated operator norm. 

This allows us to use Eq. \eqref{convergence_estimates} from Theorem \ref{newton-kantorovich-theorem} in order to assess the convergence rate $q$ which, as we recall, is defined by: 
\begin{align} 
    q&= 1 - \sqrt{1-2\eta} 
\end{align}
and is such that:
\begin{align} \label{convergence_rate_q_2}
    \lVert \tn - \bm t^* \rVert \leq \frac{q^n}{1-q} \eta,
\end{align}
We recall that $\eta$ can be calculated as:
\begin{align} \label{eta_expression}
    \eta \equiv \lVert \bm t^{(1)} - \bm t^{(0)} \rVert_1 = \sum_{\mu \in \mathcal{A}} \left| \frac{r_\mu(\bm t^{(0)})}{E_\mu(\bm t^{(0)})- E_0(\bm t^{(0)})} \right|
\end{align}
Note that if we look at \eqref{eta_expression}, we can see that
if $E_\mu(\bm{t})-E_0(\bm{t})$ is not too far from the gap between the actual ground state and the excited state corresponding to $\mu$, then one can write:
\begin{align} \label{q-homo-lumo}
    q \leq 1 - \sqrt{ 1 - 2\frac{\lVert r(\bm t^{(0)}) \rVert_1}{\Delta}}
\end{align}
where $\Delta = E_{\mathrm{es}} - E_{\mathrm{gs}}$ is the Highest Occupied Molecular Orbital - Lowest Unoccupied Molecular Orbital (HOMO-LUMO) gap of the molecule.
What we can see from Eq.~\eqref{q-homo-lumo} is that $q$ is upper-bounded by an increasing function of the ratio between the $1$-norm of $r(\bm t^{(0)})$ and the HOMO-LUMO gap of the molecule.
This also means, if we consider a fixed $\lVert r(\bm t^{(0)}) \rVert_1$, that a larger gap should lead to faster convergence.

\subsubsection{Numerical checks} \label{convergence_rate_benchmarks}
\begin{figure}[ht]
    \centering
    \includegraphics[width=0.5 \textwidth]{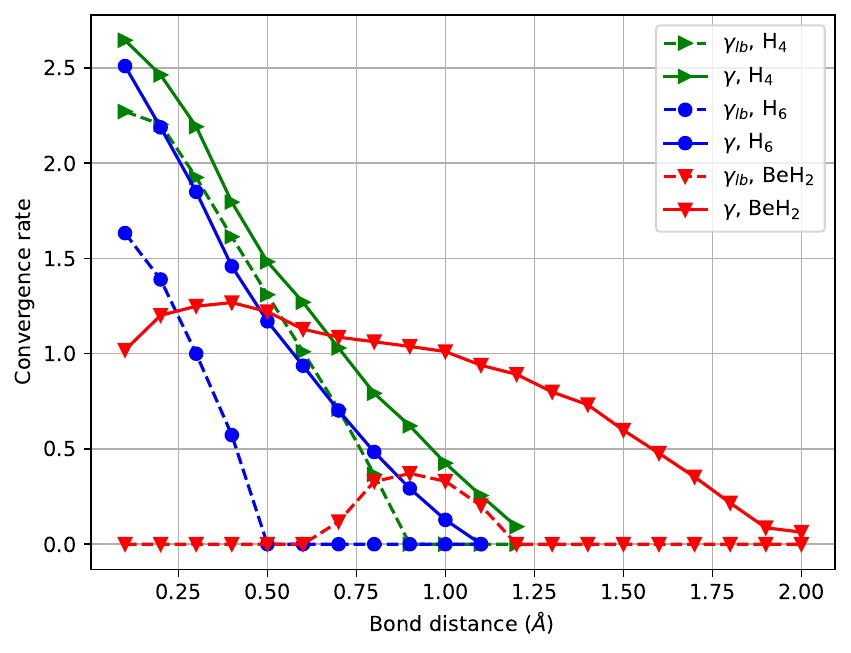}
    \caption{Actual convergence rate $\gamma$ (solid lines) and predicted convergence rate $\gamma_\mathrm{lb}$ of the approximate modified-Newton-Raphson optimizer (Eq. \eqref{modified-newton-update-rule}) in the case of $\mathrm{H_4}$ (in blue), $\mathrm{H_6}$ (in green) and $\mathrm{BeH_2}$ (in red) molecules for several bond distances.}
    \label{convergence_rates}
\end{figure}

In order to check what we have been discussing, we ran simulations of the PQE algorithm using the approximate modified Newton-Raphson optimization (Eq. \eqref{modified-newton-update-rule}). According to Eq. \eqref{convergence_rate_q_2}, the slope of $\lVert \tn - \bm{t}^{\mathrm{final}} \rVert_1$ in logplot should be close to a linear plot with a slope $- \gamma = \log(q)$. We thus fit this evolution in order to obtain the coefficient $\gamma$. We also retrieve at the beginning of the algorithm the quantity $\eta$ defined in \eqref{eq:eta_def} allowing us to obtain the lower bound of the coefficient $\gamma$:
\begin{align}
    \gamma_\mathrm{lb} = - \log(1- \sqrt{1 - 2 \eta})
\end{align}
We plot $\gamma$ and $\gamma_\mathrm{lb}$ in Figure \ref{convergence_rates} for $\mathrm{H_4}$ (green slopes), $\mathrm{H_6}$ (blue slopes) and $\mathrm{BeH_2}$ (red slopes) molecules for different bond distances.
We can see that in the case of $\mathrm{H_4}$ (green slopes), our predictions get very close to the actual value while being lower (as expected). We can however see that at larger bond distances the discrepancy becomes larger: we predict a convergence up to $0.8 \text{\AA}$ while the algorithm converges up to $1.2 \text{\text{\AA}}$. This suggests that at higher bond distances (which in the case of $\mathrm{H_4}$ means parameter vectors with larger norms) the approximation $L \approx 1$ might be an overestimation. Looking at the slopes for $\mathrm{H_6}$ (blue slopes), we still get a lower bound reproducing nicely the tendency. The predictions are however far from the true value and this again gets worse with larger bond distances, suggesting again that the true value of $L$ might be smaller than $1$. In the case of $\mathrm{BeH_2}$, although we still technically get a lower bound, we only predict convergence for a fraction of the cases where it does converge. Again, this can be explained by the fact that the value $L \approx 1$ might be too large. In the cases where we do not predict convergence, one can check that the norm of the parameter vectors gets to around $0.7$, meaning the assumptions we used to obtain $L \approx 1$ no longer hold, hence the poor predictions. Still, we can see that our derivations do not provide very accurate quantitative predictions but are still interesting for qualitative predictions.

\section{A Resource-Efficient and Robust Optimization} \label{sec:Better_opti}
\subsection{The Method}
We have seen in the previous section that it is possible to use the Newton Kantorovich Theorem in order to obtain predictions for quasi-Newton optimizations. Still, even though our hybrid update rule displays better convergence properties, it does not converge in every case, and it is based solely on the empirical observation that the standard update rule converges in many cases.

We would like to use the idea of the hybrid optimization, which consists in using a first optimization to get to a point in the manifold where we know a Newton-Raphson optimization will converge, and then using an approximate Newton-Raphson to take advantage of its faster convergence.
Previously, we used the standard optimization (Eq. \eqref{standard-update-rule}) to get to the point in question, but it does not work in every case. The first idea of our novel optimization method is the following: gradient descents on the energy have the upside of not requiring that the starting point in parameter space be close to the converged parameter.
Moreover, since we are in the end interested in the energy, making sure we reduce it each step makes sense.
Finally, one can check that (see appendix \ref{jacobian-derivations}):
\begin{align} \label{residue-gradients}
    \frac{\partial E_0}{\partial t_\mu}(\bm{t}) = 2 r_\mu(\bm{t}) + O(\lVert \bm{t}\rVert).
\end{align}
This means that instead of following the gradient at a given point to get an improvement on the energy, one can follow the residues, the upside being that we can check at each step whether a Newton-Raphson will converge, and we can still use the well-informed convergence criterion from Eq. \eqref{convergence_criterion}. This also means that as long as we are far from the solution, the residues should remain big, meaning there will not be any barren plateau or local minima phenomenon. One could argue that in cases where the gradients are small but the residues are not (which means Eq. \eqref{residue-gradients} is a poor approximation), there is no real reason that the direction given by the residue vector should yield improvements on the energy. This difficulty should be somewhat alleviated by the next additions, and if such cases arise one could always replace this step with another residue-based scheme.  

This first idea thus gives rise to the following update rule:
\begin{align} \label{first_step_better_opti}
     \tnn_\mu = \begin{cases} &\tn_\mu - 2 r_\mu(\tn) \  \text{if} \ \eta_n  > \frac1{2L} \\
     &\tn_\mu - \frac{r_\mu(\tn)}{E_\mu(\tn) -E_0(\tn)} \ \text{otherwise}
     \end{cases},
 \end{align}
where $L$ is the constant from Theorem \ref{newton-kantorovich-theorem}. 
The goal of the first update rule is thus solely to bring us to a place in the manifold where $\eta_n < \frac1{2L}$. 

Since we are primarily interested in cases where $t^*$ is far from $t^{(0)}$, we however expect the difference in direction between the gradients and the residues to be non negligible, and vanilla gradient descent with a learning rate of $1$ lacks robustness, 
we add a variable step size $\alpha_n $ such that:
\begin{align} \label{second_step_better_opti}
     \tnn_\mu = \begin{cases} &\tn_\mu - 2 \alpha_n r_\mu(\tn) \  \text{if} \  \eta_n > \frac1{2L} \\
     &\tn_\mu - \frac{r_\mu(\tn)}{E_\mu(\tn) -E_0(\tn)} \ \text{otherwise}
     \end{cases}.
 \end{align}
 $\alpha_n$ is chosen using a backtracking line search on the energy and such that it respects the Armijo condition for a parameter $c < 1$ (for more details, read \cite[sections 3.1 and 3.2]{NoceWrig06}). If such an $\alpha_n$ is not found (or rather is below a user-defined threshold $\alpha_{\mathrm{threshold}}$, such as $\alpha_{\mathrm{threshold}}=10^{-10}$), the algorithm terminates. 
 
 We could enforce the Wolfe condition \cite{NoceWrig06}, i.e add a curvature condition, but since measuring the residues is much more costly than just measuring the energy, and the residues are not really equal to the gradients, we simply choose to forego the curvature condition. Moreover, the curvature is not needed when using a backtracking line search for Newton methods.  
 
 Since this line search is fairly cheap in comparison with each optimization step, we also add a variable step length to the quasi-Newton part of the optimization in order to make it more robust. The line search is therefore performed in each case, after determining which update rule to use. We are left with the update rule:
 \begin{align} \label{third_step_better_opti}
     \tnn_\mu = \begin{cases} &\tn_\mu - 2 \alpha_n r_\mu(\tn) \  \text{if} \  \eta_n > \frac1{2L} \\
     &\tn_\mu - \alpha_n \frac{r_\mu(\tn)}{E_\mu(\tn) -E_0(\tn)} \ \text{otherwise}
     \end{cases}
 \end{align}
 The last idea of our optimization method is to take advantage of the curvature information we have gathered during the previous iterations using the BFGS update rule (Ref.~\onlinecite{NoceWrig06}). The BFGS algorithm uses curvature information from past iterations in order to iteratively construct an approximation of the Hessian of the cost function. In our case, we are interested in improving our guess of inverse Jacobian since we are using an approximation which can be fairly rough in highly correlated cases. 
 
 We therefore introduce the update rule: 
  \begin{align} \label{final_step_better_opti}
     \tnn =  \tn - \alpha_n H_n r(\tn), \  
 \end{align}
where $H_n$ will be updated using the BFGS update rule: 
 \begin{align} \label{bfgs_update}
\nonumber  H_{n+1} &= H_n + \frac{\left(s_n^T y_n + y_n^T H_n y_n \right)(s_n s_n^T)}{ \left(s_n^T y_n\right)^2}  \\
&- \frac{  H_n y_n s_n^T +s_n y_n^T H_n }{s_n^Ty_n}
 \end{align}
where $y_n = r(\tnn) - r(\tn)$ and $s_n = -\alpha_n H_n r(\tn)$.

We also use the tools we have derived so far. 
Namely, we start with $H_0 = 2I$ or $H_0 = \diag\left(\left(\frac1{E_\mu(\bm t^{(0)}) - E_0(\bm t^{(0)})}\right)_{\mu \in \mathcal{A}}\right)$  depending on the value of $\eta$ wrt. a user-defined threshold $\eta_{\mathrm{threshold}}$. Typically, we have seen that a threshold $\eta_{\mathrm{threshold}}= \frac12$ works well, but since we can see from section \ref{convergence_rate_benchmarks} that the value $L \approx 1$ can be an overestimation, and since the line search makes everything more robust, the threshold can be increased. We have found that a threshold $\eta_{\mathrm{threshold}}=1$ typically works well.
During the optimization, if we started with $H_0 = 2I$, we test at each step whether the criterion is met, and if it is at step $n$, then we use $H_{n+1} = \diag\left(\left(\frac1{E_\mu(\tn) - E_0(\tn)}\right)_{\mu \in \mathcal{A}}\right)$.  
A last improvement which is less important but makes sense, is that since we are dealing with optimizations that might go far from the starting point, and with manifolds which typically are not close to linear (otherwise we would use a quasi-Newton from the start), we choose to discard curvature information from previous iterations if the current parameter is far from the previous one. That is, if $\lVert \tnn - \tn \rVert_1 > \tau_{\mathrm{reset}} $ we use $H_{n+1} = 2I$ or  $H_{n+1} = \diag\left(\left(\frac1{E_\mu(\tnn) - E_0(\tnn)}\right)_{\mu \in \mathcal{A}}\right)$ depending on the value of $\eta_n$. $\tau_{\mathrm{reset}}$ is chosen by the user, but we have found that $\tau_{\mathrm{reset}} = \frac12$ works fairly well (notice that this value makes sense wrt. our study so far). 
 Lastly, we would like to point out that foregoing the curvature condition during the line search means there theoretically might be cases where $H_n$ is not positive-definite. Although we have not encountered this case, one can simply choose not to update $H_n$ if the update would make it non-positive-definite. Since the BFGS update rule is really not a central point of our method (its purpose is just to make our approximation of the Jacobian a bit better), it is reasonable to do so.\\
 The pseudocode of the algorithm is available under Algorithm \ref{pseudocode}.

\begin{algorithm}[H]
\caption{Our Optimization Method}
\label{pseudocode}
\begin{algorithmic}[1]
\State \textbf{Input:} Starting point $\bm t^{(0)}$, onvergence threshold $\varepsilon$, switch threshold $\eta_{\mathrm{threshold}}$ (typically $1$), reset threshold $\tau_{\mathrm{reset}}$ (typically $\frac12$), Armijo threshold $c$ (typically $10^{-4}$), linesearch stop threshold $\alpha_{\mathrm{threshold}}$ (typically $10^{-10}$). 
\State \textbf{Compute} $E_0(\bm t^{(0)})$ and the values $\left(r_\mu(\bm t^{(0)})\right)_{\mu \in \mathcal{A}}$, $\left(E_\mu(\bm t^{(0)})\right)_{\mu \in \mathcal{A}}$. \\
\State \textbf{Compute} $\eta_0 = \sum_{\mu \in \mathcal{A}} \left| \frac{r_\mu(\bm t^{(0)})}{E_\mu(\bm t^{(0)}) - E_0(\bm t^{(0)})} \right|$
\If {$\eta_0 < \eta_{\mathrm{threshold}}$} 
    \State $$H_0 = \diag\left( \frac1{\left(E_\mu(\bm t^{(0)}) - E_0(\bm t^{(0)}) \right)}\right)$$
\Else : 
     $H_0 = 2I$
    \EndIf
\While{$\epsilon_{\mathrm{T}}^{\mathcal{A}}(\tn)> \epsilon$}
    \State \textbf{Backtracking line search}: Find step size $\alpha_n$ such that: \[
        E_0\left(\tn - \alpha_n H_n r(\tn)\right) < E(\tn) -  c \cdot  \alpha_n(H_n r(\tn))^Tr(\tn)
    \]
    \If{$ \alpha <  \alpha_{\mathrm{threshold}}$} 
        \State \textbf{Return} $\tn$, $E_0(\tn)$ \EndIf
    \State \textbf{Update parameter:} $\tnn = \tn - \alpha_n H_n r(\tn)$
    \State \textbf{Compute} $r(\tnn), \left(E_\mu(\tnn) \right)_{\mu \in \mathcal A}$
    \State \textbf{Compute} $\eta_{n+1}$
    \State \textbf{Compute} 
        $$\epsilon_\mathrm{T}^\mathcal{A}(\tnn) = \sum_{\mu \in \mathcal{A}} \frac{r_\mu(\tnn)^2}{E_0(\bm t^{(0)}) - E_0(\tnn)}$$
    \If{$\eta_{n+1} < \eta_{\mathrm{threshold}}$ and $\eta_{n} > \eta_{\mathrm{threshold}} $} 
    
        \State \textbf{Set}  $H_{n+1} = \diag\left( \frac1{\left(E_\mu(\tnn) - E_0(\tnn) \right)}\right)$
    \Else   
        \If{$ \lVert \tnn - \tn\rVert_1 > \tau_{\mathrm{reset}}$}
        \If{$\eta_{n+1} < \eta_{\mathrm{threshold}}$} 
            \State \textbf{Set} $$ \qquad H_{n+1} = \diag\left( \frac1{\left(E_\mu(\tnn) - E_0(\tnn) \right)}\right)$$
        \Else 
            \State \textbf{Set} $$H_{n+1} = 2I $$
        \EndIf
        \Else 
            \State \textbf{Update inverse Jacobian guess:} Eq. \eqref{bfgs_update}
        \EndIf
        
    \EndIf
\EndWhile
\State \textbf{Return} $\tnn$, $E_0(\tnn)$
\end{algorithmic}
\end{algorithm}

 \subsection{Benchmarks} 
   \begin{figure*}[t]
     \centering
     \includegraphics[width=1\textwidth]{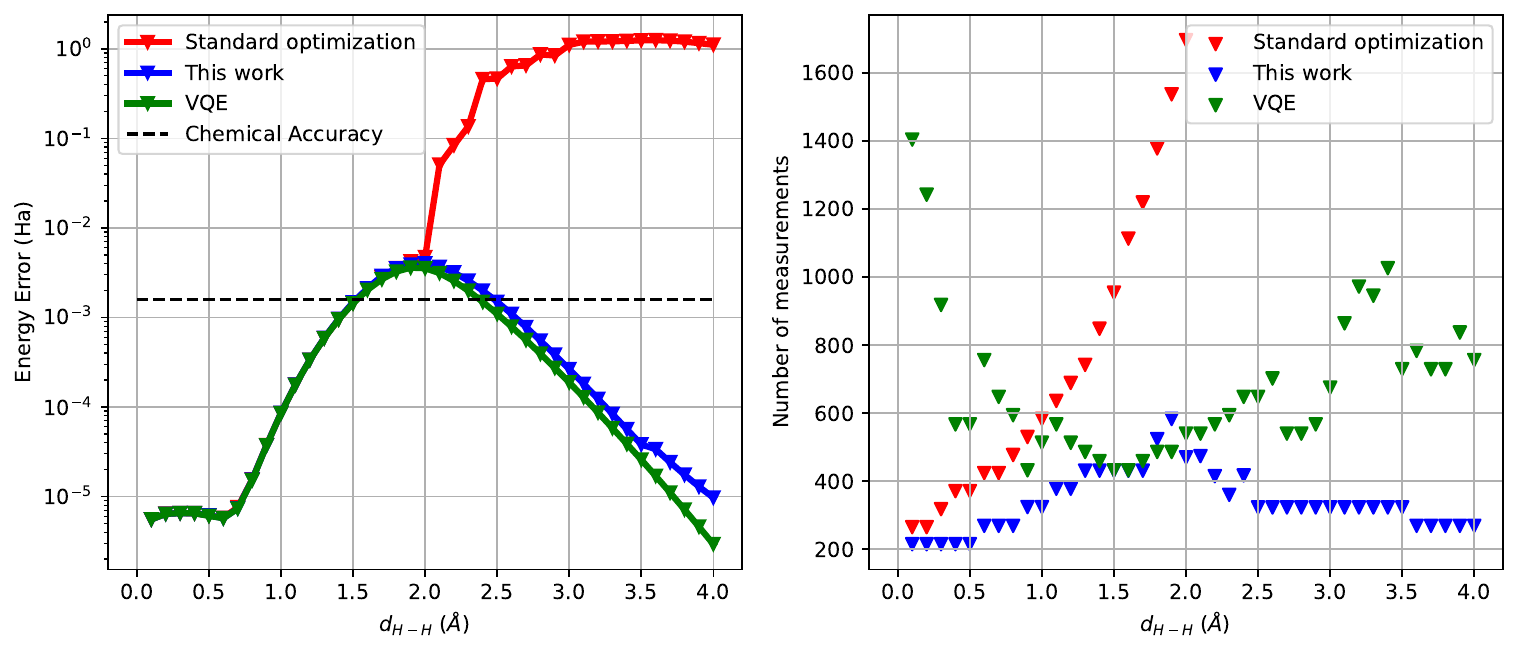}
     \caption{Same as Fig. \ref{BeH2_vqe_sn_truc_1} but on an $\mathrm{H_4}$ molecule}
     \label{H4_vqpqe}
 \end{figure*}
   \begin{figure*}[t]
     \centering
     \includegraphics[width=1\textwidth]{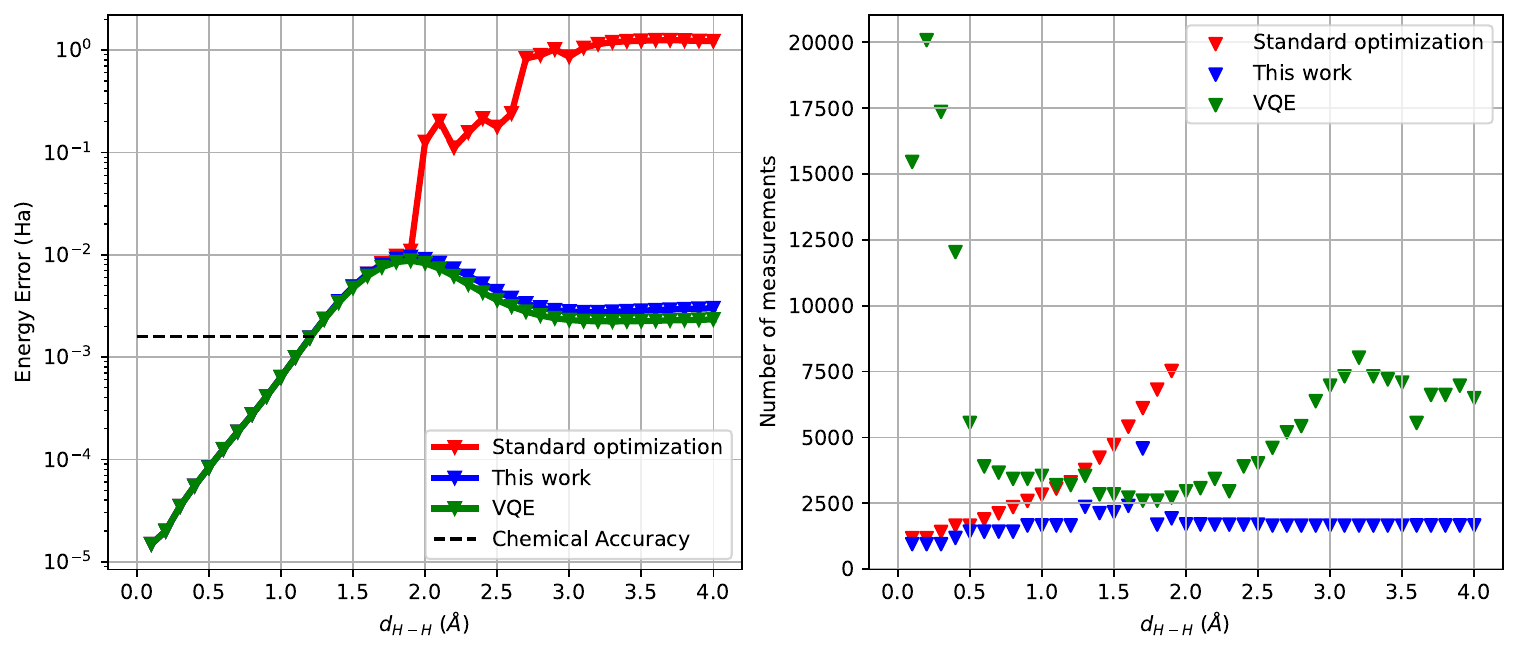}
     \caption{Same as Fig. \ref{BeH2_vqe_sn_truc_1} but on an $\mathrm{H_6}$ molecule}
     \label{H6_vqepqe}
 \end{figure*}
   \begin{figure*}[t]
     \centering
     \includegraphics[width=1.1\textwidth]{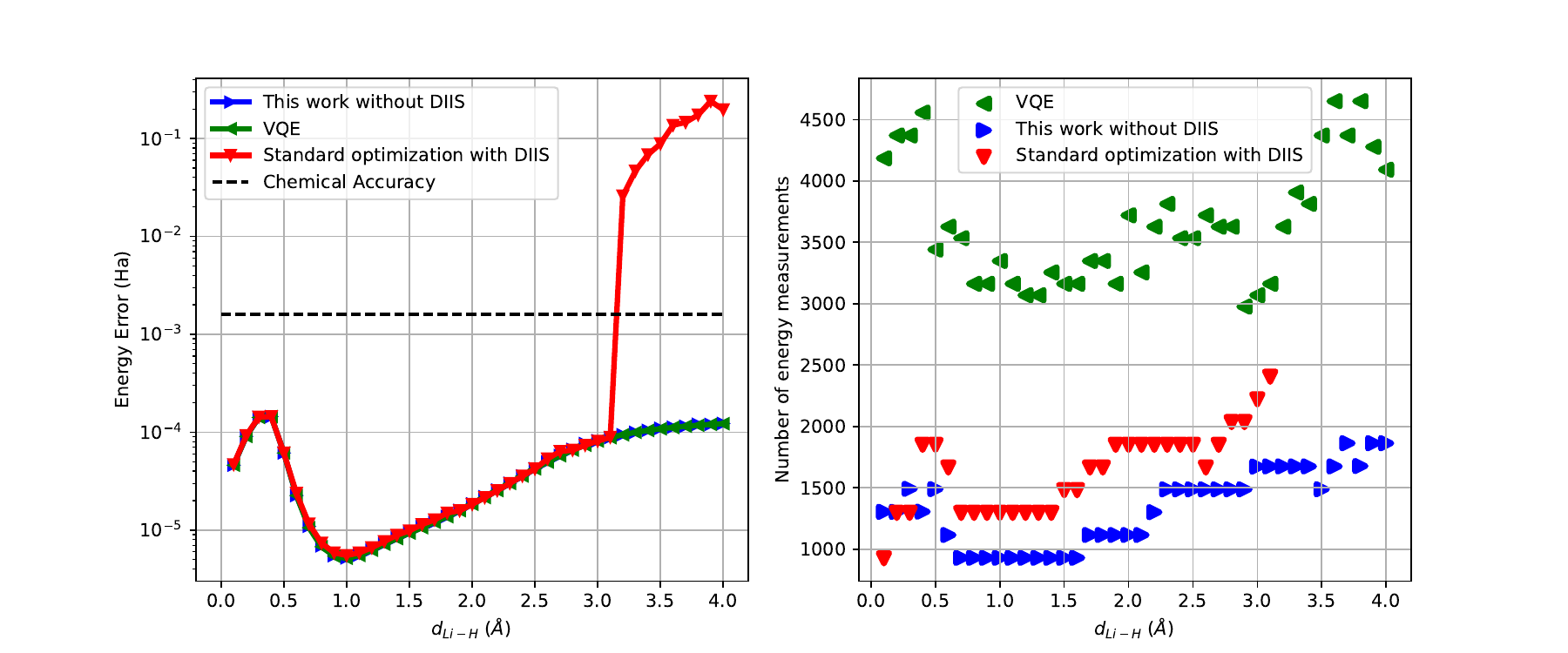}
     \caption{Same as Fig. \ref{BeH2_vqe_sn_truc_1} on an $\mathrm{LiH}$ molecule, with DIIS added to the standard optimization}
     \label{LIH_vqepqe_diis}
 \end{figure*}
 
 In order to test our optimization, we ran our implementation of it on Eviden's Qaptiva platform, for $\mathrm{H}_4$, $\mathrm{H}_6$, $\mathrm{LiH}$ and $\mathrm{BeH}_2$ molecules using the same ansatz as mentioned in subsection \ref{benchmarks_hybrid}.
 We used the same threshold $\varepsilon = 10^{-5}$ for both the standard optimization and our optimization.
 We also ran our implementation of the variational quantum eigensolver, again with the same ansatz, where the parameters $\bm t$ were optimized using the implementation of BFGS provided by ScipyMinimize with a threshold for the gradient norm of $10^{-5}$. In order to compare the algorithms, we plot the difference between the converged energy and the energy computed using FCI (ran using PySCF) which we dub the energy error, and we plot the total number of energy measurements used by the algorithm. The results can be found in Figures \ref{BeH2_vqe_sn_truc_1}, \ref{H6_vqepqe} and \ref{H4_vqpqe}. 
 Again, we wish to compare the performance of the update rules and thus did not include DIIS (except for Fig. \ref{LIH_vqepqe_diis}).
 
We can see when looking at Fig. \ref{BeH2_vqe_sn_truc_1} that PQE (both using the standard optimization and our optimization) performs comparably to VQE in terms of energy error for bond distances smaller than $3$ \AA. Beyond that, the standard optimization stops converging while ours still does. The results obtained are still quite comparable to VQE, although there is a slight discrepancy, the origin of which has been discussed in subsection \ref{criterion_benchmarks}. In terms of number of measurements however, we can see that our algorithm beats both VQE and the standard implementation of PQE. 
 
The same can be said when looking at $\mathrm{LiH}$ (Fig. \ref{BeH2_vqe_sn_truc_1}): PQE with our optimization yields results comparable with that of VQE, but does so with far less measurements than both VQE and the standard implementation of PQE, which does not converge past $3$ \AA. 
 
In the cases of Hydrogen chains $\mathrm{H_4}$ and $\mathrm{H_6}$ (Figures \ref{H4_vqpqe} and \ref{H6_vqepqe}), the standard implementation of PQE quickly gets worse than VQE in terms of number of energy measurements, and stops converging for bond distances larger than $2$ \AA. Our optimization however, converges in every case, and yields comparable results to VQE accuracy-wise, while being mostly superior measurements-wise. The cases where our algorithm takes a bit more measurements correspond to cases where the global minimum of the energy is in a place of the landscape where the convergence criterion is not met, therefore the algorithm performs a few superfluous steps without being able to get much improvement on the residual norm, and then stops when the line search yields steps too small. 

 Finally, looking at Fig. \ref{LIH_vqepqe_diis}, we can see that our update rule performs better than the standard update rule with DIIS. DIIS could be added to our optimization, but an efficient integration is less straightforward (the Jacobian used can change significantly between iterations) and thus outside the scope of this study.
Overall, our method yields results comparable to the ones given by VQE, while requiring in most cases far fewer energy measurements and offering the theoretical guarantees we have previously discussed.

\section{Conclusion}

In this work, we have shown evidence of promising aspects of residue-based optimizations for unitary coupled cluster calculations on quantum processors.
These promising aspects include robust convergence properties when getting close enough to the solution.
These convergence features are similar to the ones which can be obtained with the variance-based variational quantum eigensolver \cite{zhang2020variationalquantumeigensolversvariance}, albeit without the additional computational overhead.
We anticipate that these convergence properties could potentially at least partially solve some of the challenges encountered with the standard VQE approach, such as barren plateaus and local minima.
We have also shown, using our novel optimization method, that such methods can more than rival the variational quantum eigensolver in terms of computational cost.
However, our method still faces challenges in the same capacity as VQE, mainly measurement overhead and noise sensitivity for NISQ computations, which means that even with better hardware, more algorithmic improvements will be needed in order to obtain a viable alternative to classically computed coupled cluster, or an efficient state preparation method for quantum phase estimation.

\section*{Acknowledgements}
This work was supported by the ANR Research Collaborative Project 'QuRes' (Grant ANR-PRC-CES47-0019). As part of the MetriQs-France program, this work was supported by France 2030 under the French National Research Agency grant number ANR-22-QMET-0002. As part of the HQI initiative, this work was supported by France 2030 under the French National Research Agency award number ANR-22-PNCQ-0002. The authors thank Alexia Auffèves for her involvement in making this work possible.
The authors also acknowledge useful discussions with Cyrielle Martin, Mohammad Hassan, Yvon Maday, Robert Whitney and Victor Kadri.
The computations were performed on the Eviden Qaptiva platform.

\appendix
\section{Mathematical derivations}
\subsection{Jacobian map approximations} \label{jacobian-derivations}
In this subsection, we present the calculations and approximations allowing us to obtain the approximate Newton-Raphson and approximate modified-Newton-Raphson update rules (Eq. \eqref{approximate_newton} and \eqref{approximate_modified}). The main assumption here will be that the Hartree Fock approximation is already a good step towards computing the ground state.
This will lead to our two main assumptions. The first one is the assumption that the Hamiltonian written in the Hartree Fock basis (molecular orbitals) is already close to being diagonal, and that it will remain so on the rotated basis (i.e, transformed by $\U(\bm{t})$) during the optimization.
Mathematically speaking, we consider:
\begin{align}
    |\bra{\Phi_\mu} \U^\dag(\bm{t}) \H \U(\bm{t}) \ket{\Phi_\nu}| \ll 1
\end{align}
for all $\bm{t}$ and $\nu \neq \mu$. Note that this implies that the quasi-Newton update rules we derive will be of order $1$ in off-diagonal coefficients of the Hamiltonian.
We will also assume the solution $\bm{t}^*$ of the PQE equations to be small in norm.

In what follows, we suppose a fixed parametric ansatz $\bm{t} \rightarrow \U(\bm{t}) =\prod_{\mu\in \mathcal{A}} e^{t_\mu \hat{\kappa}_\mu}$ and introduce a partial order on $\mathcal{A}$: $\mu > \nu$ iff $\U = \cdots e^{t_\nu \hat{\kappa}_\nu} \cdots e^{t_\mu \hat{\kappa}_\mu} \cdots$. 
We also introduce the notation $\bm{t} \rightarrow \Ubar_\nu(\bm{t}) = \prod_{\nu' > \nu}  e^{t_\nu \hat{\kappa}_\nu} $. 
It will be assumed that indices $\mu, \nu$ are in $\mathcal{A}$ unless specified otherwise. 

We recall the Baker-Campbell-Hausdorff (BCH) expansion: given two operators $X$ and $Y$, one can write: 
\begin{align} \label{BCH}
    e^Y X e^{-Y} = X + \left[X, Y\right] + \frac12 \left[\left[X, Y\right], Y\right] + \cdots
\end{align}
In order to obtain a general formula for the Jacobian map $\bm{t}\rightarrow J(\bm{t})$ of the residue map, we expand $r_\mu(\bm{t} + \Delta \bm{t})$ and identify the coefficients of $J(\bm t)$ as the coefficients of the first order terms.
Thus we expand $\U(t + \Delta t)^\dag \H \U(t + \Delta t)$ to first order in $\Delta t$ using BCH. One can easily check that we get:
\begin{align} \label{general_jacobian_formula}
    J_{\mu, \nu}(\bm{t}) = \bra{\Phi_\mu} \left[ \U(\bm{t})^\dag \H \U(\bm{t}), \Ubar_\nu (\bm{t})^\dag \kap_\nu \Ubar_\nu(\bm{t}) \right] \ket{\Phi_0}
\end{align}
For $t = 0$, we simply get:
\begin{align} \label{general_jacobian_formula_in_zero}
J_{\mu, \nu}(0) = \bra{\Phi_\mu} \left[ \H , \kap_\nu\right] \ket{\Phi_0}.
\end{align}
If, in addition, we further approximate $\H$ with its diagonal, we get:
\begin{align} \label{j-zero-diag-approximation}
    J(0)_{\mu, \nu} = \delta_{\mu, \nu} \left(E_\mu(0) - E_0(0)\right),
\end{align}
with $E_\mu(0)$ and $E_0(0)$ defined in the main text, Eq.~\eqref{eq:E_mu}.
This approximation is what yields the approximate modified-Newton-Raphson rule introduced in Eq. \eqref{approximate_modified}. 

Another way to approximate Eq. \eqref{general_jacobian_formula_in_zero} is to consider that $\H$ is composed of the Hartree Fock Hamiltonian $H_{HF} = \sum_{p} \epsilon_p \hat{a}_p^\dag  \hat{a}_p $ and a perturbation $\hat{F}$ containing all the correlation terms of $\H$. This corresponds in Eq. \eqref{hamiltonian} to $h_{pq} = \delta_{p,q}\epsilon_p$

This yields the approximation:
\begin{align} \label{j-zero-diag-approximation-hf}
    J(0)_{\mu, \nu} = - \delta_{\mu, \nu}\Delta_\mu,
\end{align}
where $\Delta_\mu = \epsilon_i +\epsilon_i + \cdots - \epsilon_a - \epsilon_b - \cdots $ is the standard M\o ller Plesset denominator corresponding to the excitation $\mu$. This approximation is what leads to the update rule in Eq. \eqref{standard-update-rule} which is introduced in \cite{Stair_2021}. 

Now going back to $J(\bm{t})$, using Eq. \eqref{general_jacobian_formula} and approximating $\U(\bm{t})^\dag\H \U(\bm{t})$ with its diagonal $\diag\left(E_\mu(\bm{t})\right)$ we get:
\begin{align} \label{jacobian_diag_h}
    J_{\mu, \nu}(\bm{t}) = (E_\mu(\bm{t}) -E_0(\bm{t}))\bra{\Phi_\mu} \Ubar_\nu^\dag(\bm{t}) \kap_\nu\Ubar_\nu(\bm{t}) \ket{\Phi_0}.
\end{align}
Looking at the case $\nu \neq \mu$, we can expand $\bra{\Phi_\mu} \Ubar_\nu^\dag(\bm{t}) \kap_\nu\Ubar_\nu(\bm{t}) \ket{\Phi_0}$ using BCH:
\begin{align} \label{first_order_u_kap_u}
\nonumber \bra{\Phi_\mu} \Ubar_\nu^\dag(\bm{t}) \kap_\nu\Ubar_\nu(\bm{t}) \ket{\Phi_0} = \bra{\Phi_\mu}\kap_\nu\ket{\Phi_0} +  \\ \qquad \sum_{\nu'>\nu}t_{\nu'} \bra{\Phi_\mu} \left[\kap_\nu, \kap_{\nu'}\right]\ket{\Phi_0}  + o(\lVert \bm{t}\rVert),
\end{align}
where the term $o(\lVert \bm{t}\rVert)$ also contains higher-order commutators.
When $\nu \neq \mu$ we have $\bra{\Phi_\mu}\kap_\nu\ket{\Phi_0} = 0$. We then choose to neglect the other terms, considering that in cases with a low correlation the Hartree Fock state will be good enough to assume $\lVert t \rVert_1 \ll1$, and that even in more correlated cases most of the commutators will be zero. 
If $\mu = \nu$, then we get $\bra{\Phi_\mu}\kap_\nu\ket{\Phi_0} = 1$, and we will neglect the other terms of the expansion for the aforementioned reasons. 

We end up with the following approximation:
\begin{align} \label{jacobian_diag_approx}
    J_{\mu, \nu}(\bm{t}) = \delta_{\mu, \nu}\left(E_\mu(\bm{t}) - E_0(\bm{t})\right),
\end{align}
which allows us to derive the update rule introduced in Eq. \eqref{approximate_newton}.

\subsection{The constant $L$}
The goal here is to find $L$ such that $\lVert J^{-1}\left( J(\bm{t}) - J(\bm t^{(0)})\right) \rVert \leq L \lVert \bm{t}\rVert $ (where $\bm t^{(0)}$ is not necessarily $0$ since we sometimes perform a few iterations from another method beforehand) so that we can use the Newton-Kantorovich Theorem to predict the convergence properties of the algorithm.
We first go back to Eq. \eqref{jacobian_diag_h}. Our goal is to use its first order in $t$ expansion. We first notice that using BCH we get:
\begin{align}
    E_\mu(\bm{t}) = E_\mu(0) + \sum_{\nu \in \mathcal{A}} t_\nu \bra{\Phi_\mu} \left[H, \kap_\nu \right]\ket{\Phi_\mu} + o(\lVert \bm{t}\rVert),
\end{align}
which within our diagonal approximation of $\H$ can be further approximated as:
\begin{align}
E_\mu(\bm{t}) = E_\mu(0) + o(\lVert \bm{t}\rVert),
\end{align}
and the same can be said of $E_0(\bm{t})$.
We will from now on omit the term $o(\lVert \bm{t}\rVert)$. 

Now, since we are interested in deriving an expression for $\left(J(\bm t^{(0)})^{-1}\left(J(\bm{t}) - J(t^{(0)})\right)\right)_{\mu,\nu} $ where $\bm t^{(0)}$ is the starting point of the quasi-Newton optimization, we need to have first an expression in first order of $\lVert \bm t^{(0)} \rVert $ of $J(\bm t^{(0)})^{-1}$. 

Using \eqref{first_order_u_kap_u} and \eqref{jacobian_diag_h}, we can see that to first order: 
\begin{align}
    J(\bm t^{(0)}) = D(I + S)
\end{align} where $D = \diag\left( E_\mu(\bm t^{(0)}) - E_0(\bm t^{(0)})\right)_{\mu, \nu}$ and $S = \left( \sum_{\nu'>\nu}t^{(0)}_{\nu'} \bra{\Phi_\mu} \left[\kap_\nu, \kap_{\nu'}\right]\ket{\Phi_0}\right)_{\mu, \nu}$.
Thus, we get that in first order, the inverse of $J(\bm t^{(0)})$ reads:
\begin{align} \label{eq-j-inverse}
J(\bm t^{(0)})^{-1} = (I - S)D^{-1}
\end{align}
One can thus verify that in first order in both $\bm t$ and $\bm t^{(0)}$ we get:
\begin{align}
    J(\bm t^{(0)})^{-1} J(\bm t)_{\mu, \nu} = \delta_{\mu,\nu} + \sum_{\nu'>\nu} (t_{\nu'}- t^{(0)}_{\nu'}) \bra{\Phi_\mu} \left[\kap_\nu, \kap_{\nu'}\right]\ket{\Phi_0}
\end{align}
Thus:
\begin{align}
    \nonumber \left(J(\bm t^{(0)})^{-1}\left(J(\bm{t}) - J(\bm t^{(0)}) \right)\right)_{\mu,\nu}  &=  \\ 
     \sum_{\nu' >\nu } (t_{\nu'}- t^{(0)}_{\nu'})&\bra{\Phi_\mu} \left[\kap_\nu, \kap_{\nu'}\right] \ket{\Phi_0}
\end{align}
Since the parameter space is of finite dimension, we could choose any norm to determine $L$ but it is best to use the norm for which the constant $L$ in the end is the smallest. In what follows we use for simplicity the $1$-norm $\lVert \cdot \rVert_1$ of $\mathbb{R}^N$, the associated operator norm of which is the $1$-norm taken wrt. the columns, which we will also write $\lVert \cdot \rVert_1$. We get:
\begin{align}
   \nonumber \lVert J(\bm t^{(0)})^{-1}&(J(\bm{t})-J(\bm t^{(0)})) \rVert_1 = \\ &\sup_{\nu} \sum_{\mu \in \mathcal{A}}  \left| \sum_{\nu' > \nu}(t_{\nu'}- t^{(0)}_{\nu'})'\bra{\Phi_\mu}\left[ \kap_\nu, \kap_{\nu'} \right] \ket{\Phi_0}\right| \label{jacobian_norm}\\
   &\leq \sup_{\nu}  \sum_{\nu' > \nu} |t_{\nu'}- t^{(0)}_{\nu'}| \sum_{\mu \in \mathcal{A}} \left|\bra{\Phi_\mu}\left[ \kap_\nu, \kap_{\nu'} \right] \ket{\Phi_0}\right|
\end{align}

As discussed earlier, most of the commutators in the sum will be zero. To get a rough upper bound, one can notice that for each $(\nu, \nu')$ there is at most one $\mu$ such that $\kap_\nu$ and $\kap_{\nu'}$ do not commute, and either $\kap_\nu\kap_{\nu'}\ket{\Phi_0} = \pm \ket{\Phi_\mu} $ or $\kap_{\nu'}\kap_\nu\ket{\Phi_0} = \ket{\Phi_\mu} $. Therefore we simply get:\begin{align}
    \lVert J(\bm t^{(0)})^{-1}(J(\bm{t})-J(\bm t^{(0)})) \rVert_1 \leq \lVert \bm{t}\rVert_1,
\end{align} 
and thus $L =1$. 

This is the tightest upper bound on Eq. \eqref{jacobian_norm} if we consider $\bm t $ to be any point in a ball around $\bm t^{(0)}$ , but in general since many components of $\bm t$ are nonzero it is a somewhat rough upper bound, because for each $ \nu$ not every $\nu'>\nu$ will result in a nonzero term $\sum_{\mu} \left|\bra{\Phi_\mu}\left[ \kap_\nu, \kap_{\nu'} \right] \ket{\Phi_0}\right|$.

\subsection{About gradients}
We are interested in establishing the relation:
\begin{align} \label{residues_gradients}
    \frac{\partial E_0}{\partial t_\mu}(\bm{t}) = 2 r_\mu(\bm{t}) + O(\lVert \bm{t}\rVert)
\end{align}
One can easily check that using the notations we have previously introduced, we have:
\begin{align}
     \frac{\partial E_0}{\partial t_\mu}(\bm{t}) =  \bra{\Phi_0} \left[ \U(\bm{t})^\dag \H \U(\bm{t}), \Ubar_\nu (\bm{t})^\dag \kap_\mu \Ubar_\nu(\bm{t}) \right] \ket{\Phi_0},
\end{align}
which when using the value at $\bm t = 0$ leaves us with:
\begin{align}
     \frac{\partial E_0}{\partial t_\mu}(\bm{t}) &=  \bra{\Phi_0} \left[ \H,\kap_\mu  \right] \ket{\Phi_0} + O(\lVert \bm{t} \rVert) \\
     &= 2 r_\mu(\bm{t}) + O(\lVert \bm{t} \rVert)   
\end{align}
\bibliography{biblio}
\end{document}